\documentclass{JFM-FLM_Au}

\usepackage{physics}
\usepackage{caption} 
\usepackage{subcaption} 
\usepackage{ragged2e} 
\usepackage{booktabs} 
\usepackage{multirow}

\usepackage{amsmath}
\allowdisplaybreaks

\usepackage{graphicx}

\usepackage{hyperref}
\hypersetup{
    colorlinks = true,
    urlcolor   = blue,
    citecolor  = blue,
    linkcolor  = blue,
}

\lefttitle{Thota Srinivas}
\righttitle{Journal of Fluid Mechanics}

\title{Bifurcation analysis of particulate Rayleigh-B\'enard convection}

\author{Thota Srinivas\aff{1}}

\affiliation{\aff{1}International Centre for Theoretical Sciences, Tata Institute of Fundamental Research, Bengaluru, 560089, Karnataka, India}

\corresau{Thota Srinivas, thota.srinivas@icts.res.in}

\begin{document}
\maketitle

\begin{abstract}
Particles carried through a convecting fluid layer exchange momentum and heat with the fluid, and so change not only where convection starts but also how it saturates. We study this coupling in the particulate Rayleigh--B\'enard system, in which a dilute stream of thermally inertial particles is injected at one wall and removed at the other. An Eulerian two-fluid model retaining drag, pressure-gradient, added-mass, and buoyancy forces and interphase heat exchange describes heavy particles and bubbles through a single density parameter $\beta$. After reproducing the linear stability results of \citet{raza2025impact}, we use a multiple-scale expansion and a Fredholm solvability condition to compute the first Landau coefficient $a_1$ near onset. At low and moderate particle loading, the bifurcation is supercritical for every density ratio considered, extending the conclusion of \citet{srinivas2025weakly} for very heavy particles. At higher flux, however, $\Re{a_1}$ changes sign and the onset becomes subcritical over a broad band of density ratios, from moderately heavy particles injected below their terminal speed to most light particles. In the plane of density ratio and injection velocity, this band is bounded by lines of tricritical points. Splitting $a_1$ by the balance equation shows that its sign is set almost entirely by the fluid energy equation, while the direct nonlinear terms of the particle equations are five to six orders of magnitude smaller. The particles thus change the nature of the bifurcation indirectly, through the base state and the linear mode they help to shape, rather than through their own nonlinear dynamics.
\end{abstract}

\begin{keywords}
B\'enard convection, particle/fluid flow, bifurcation
\end{keywords}


\section{Introduction}\label{sec:intro}
A horizontal layer of fluid heated from below is perhaps the most studied example of a hydrodynamic instability. Since the work of \citet{rayleigh1916convection} it has been known that the quiescent conducting state loses stability once the Rayleigh number exceeds a critical value, which for two rigid, isothermal walls is $Ra_c\approx1708$ \citep{chandrasekhar1961hydrodynamic}. What happens just beyond this threshold was settled by the weakly nonlinear theories of \citet{malkus1958finite}, \citet{stuart1960non}, \citet{watson1960non} and \citet{schluter1965stability}. In the Boussinesq limit the bifurcation to two-dimensional rolls is a supercritical pitchfork: the amplitude of convection grows continuously from zero as the square root of the distance from onset, and nothing is observed below $Ra_c$. This picture is robust, but it is not universal. Effects that break the up--down symmetry of the layer, such as the temperature dependence of the fluid properties, introduce quadratic interactions that can make the onset of hexagonal convection subcritical \citep{busse1978non}. Reviews of the pattern-forming and turbulent regimes are given by \citet{bodenschatz2000recent} and \citet{ahlers2009heat}.

In many settings the convecting fluid is not pure. Water droplets and ice crystals in clouds, ash in volcanic plumes, sediment in lakes and estuaries, bubbles in boiling and electrolytic cells, and the solid particles used as heat carriers in solar receivers all move through a thermally stratified fluid while exchanging momentum and heat with it. The dispersed phase can modify the flow appreciably even at small volume fractions \citep{balachandar2010turbulent}. For thermal convection, this has been studied mostly by simulation. \citet{oresta2013effects} considered settling, thermally coupled point particles in weakly turbulent Rayleigh--B\'enard convection, \citet{park2018rayleigh} separated the roles of momentum and thermal coupling, \citet{lakkaraju2013heat} examined how vapour bubbles change the heat transport, and \citet{chen2024particle} resolved the particles directly. These studies describe developed convection; they say little about how convection first appears.

The onset problem was posed in a clean form by \citet{prakhar2021linear}. In their particulate Rayleigh--B\'enard (pRB) system, particles much denser than the fluid are injected uniformly through the upper wall at a prescribed velocity and temperature and leave through the lower wall, so that the base state is steady and horizontally uniform. Using a two-fluid model, they found that the particles always delay the onset of convection. \citet{raza2024stabilization} extended the model to particles of arbitrary density, including bubbles, by adding the pressure-gradient and added-mass forces of the Maxey--Riley equation \citep{maxey1983equation, auton1988force} through the density parameter $\beta$ commonly used for inertial particles and bubbles \citep{calzavarini2008dimensionality}. They showed that both heavy and light particles stabilise the layer. \citet{raza2025impact} then examined the role of the injection velocity. Injecting heavy particles below their terminal speed stabilises the system further, while slow injection of light particles destabilises it; for bubbles, the critical Rayleigh number can even become negative, meaning that the layer convects when heated from above. Thermal inertia of the particles adds a further stabilising effect \citep{raza2025thermal}.

Linear theory identifies the threshold but not the type of bifurcation, and the type matters. If the bifurcation is subcritical, finite-amplitude convection can be sustained below $Ra_c$, the transition is hysteretic, and the linear threshold is no longer a reliable guide to when convection will be seen. In a previous study \citep{srinivas2025weakly} we carried out a weakly nonlinear analysis of particle-laden convection for water droplets in air, a case in which the particles are about a thousand times denser than the fluid, so that drag and gravity are the only forces that matter. There the bifurcation was always supercritical. It is not obvious that this conclusion survives once the density contrast is reduced. For light particles the Archimedes force on the dispersed phase is no longer negligible; the pressure-gradient and added-mass forces enter the momentum balance, and the results of \citet{raza2025impact} show that particles alone can drive the instability. Each of these effects can feed into the nonlinear saturation of the convective mode, and there is no a priori reason why they should all act in the stabilising direction.

The aim of this paper is to determine the nature of the bifurcation in the pRB system over the full range of density ratios, from very heavy particles to bubbles, and to identify where the bifurcation changes from supercritical to subcritical. We use the two-fluid model of \citet{raza2024stabilization, raza2025impact} and extend the multiple-scale analysis of \citet{srinivas2025weakly} to include the additional forces. The governing equations and the base state are presented in \S~\ref{sec:full equations} and \S~\ref{sec:linear}, where we also summarise the linear stability results that the weakly nonlinear theory builds on. The derivation of the cubic Landau equation is outlined in \S~\ref{sec:finite amplitude}, with the details collected in the appendices. The results are discussed in \S~\ref{sec:results}, and the main conclusions are drawn in \S~\ref{sec:conclusions}.

\section{Problem definition and the governing equations}\label{sec:full equations}
\begin{gather}\label{eq:fluid mass balance}
    \nabla\cdot\boldsymbol{u}=0,\\
    d_t\alpha=-\alpha\,\qty(\nabla\cdot\boldsymbol{w}),\\
    D_t\boldsymbol{u}=-\frac{1}{\rho}\,\nabla p +\nu\,\nabla^2\boldsymbol{u}+\qty[1-\beta_T\,\qty(T-T_r)]\boldsymbol{g}+\alpha\qty[\qty(D_t\boldsymbol{u}-\boldsymbol{g})+\frac{\rho_p}{\rho}\,\qty(\boldsymbol{g}-d_t\boldsymbol{w})],\\
    d_t\boldsymbol{w}=\beta \,D_t\boldsymbol{u} +\frac{\boldsymbol{u}-\boldsymbol{w}}{\tau_p}+\qty(1-\beta)\,\boldsymbol{g},\\
    D_tT=\kappa\,\nabla^2T+\alpha\qty[D_tT-E\frac{T-T_p}{\tau_T}],\\
    d_tT_p=\frac{T-T_p}{\tau_T},
\end{gather}
for all $\qty(\boldsymbol{x},\,t)\in\qty{\mathbb{R}\times\qty(-H/2,\, H/2)\times\qty(0,\, \infty)}$ with position vector $\boldsymbol{x}=x\,\vu{e}_x+z\,\vu{e}_z$. Where $\boldsymbol{u}=u_x\,\vu{e}_x+u_z\,\vu{e}_z$ is the fluid velocity vector with $u_x$ and $u_z$ being the horizontal and vertical components. Similarly, the particulate velocity field is represented by $\boldsymbol{w}=w_x\,\vu{e}_x+w_z\,\vu{e}_z$. The time scales $\tau_p=d_p^2/(12\beta\nu)$ and $\tau_T=Ed_p^2/(12\kappa)$ represent the momentum and thermal relaxation time scales, with particle size $d_p$, kinematic viscosity $\nu$, thermal diffusion coefficient $\kappa$, density ratio parameter $\beta$ and heat capacity ratio $E$. The time derivative operators $D_t()=\partial_t()+\boldsymbol{u}\cdot\nabla()$ and $d_t()=\partial_t()+\boldsymbol{w}\cdot\nabla()$ are the material derivatives following the fluid and particle path lines, respectively. The gradient operator is defined as $\nabla()=\vu{e}_x\,\partial_x()+\vu{e}_z\,\partial_z()$.

\subsection{Non-dimensional equations}
The length is taken as the distance between the hot and cold surfaces $H$, and the time scale is taken as the thermal diffusion time scale $\mathcal{T}=H^2/\kappa$. The non-dimensional variables are defined in the following form:
\begin{align}
    \qty(X,\, Z,\, \mathcal{T})=\qty(\frac{x}{H},\, \frac{z}{H},\, \frac{t}{H^2/\kappa}), \quad \qty(\boldsymbol{U},\,\boldsymbol{W})=\qty(\frac{\boldsymbol{u}}{\qty(\kappa/H)},\, \frac{\boldsymbol{w}}{\qty(\kappa/H)}),\\
    P=\frac{p}{\rho\qty(\kappa^2/H^2)}+GaPr^2Z,\quad  (\Theta,\, \Theta_p)=\qty(\frac{T-T_r}{\Delta T},\,\frac{T_p-T_r}{\Delta T}).
\end{align}
The resulting non-dimensional equations are given by
\begin{gather}\label{eq:nondim fluid mass balance}
    \nabla\cdot\boldsymbol{U}=0,\\
    d_\mathcal{T}\alpha=-\alpha\qty(\nabla\cdot\boldsymbol{W}),\\ \nonumber
    D_\mathcal{T}\boldsymbol{U}=-\nabla P+Pr\nabla^2\boldsymbol{U}+RaPr\,\Theta\, \vu{e}_z\\
    +\frac{\alpha}{2}\qty[\qty(\beta-1)\qty(D_\mathcal{T}\boldsymbol{U}+GaPr^2\,\vu{e}_z)-12Pr\,\qty(3-\beta)\Phi^{-2}\qty(\boldsymbol{U}-\boldsymbol{W})],\\
    d_\mathcal{T}\boldsymbol{W}=\beta D_\mathcal{T}\boldsymbol{U} + {12Pr\beta}\,{\Phi^{-2}}\qty(\boldsymbol{U}-\boldsymbol{W}) - \qty(1-\beta)GaPr^2\,\vu{e}_z,\\
    D_\mathcal{T}\Theta=\nabla^2\Theta+\alpha\qty[D_\mathcal{T}\Theta-12\Phi^{-2}\qty({\Theta-\Theta_p})],\\ \label{eq:nondim particle energy balance}
    d_\mathcal{T}\Theta_p=12E^{-1}\Phi^{-2}\qty(\Theta-\Theta_p)
\end{gather}
for all $\qty(\boldsymbol{X},\, \mathcal{T})\in\qty{\mathbb{R}\times\qty(-1/2,\, 1/2)\times\qty(0,\, \infty)}$ with position vector $\boldsymbol{X}=X\,\vu{e}_x+Z\,\vu{e}_z$. The boundary conditions are given by
\begin{equation}\label{eq:full BCs}
    \left.\begin{array}{l}
    \mbox{At }Z=1/2:\,\boldsymbol{U}=0 \mbox{ and } \Theta=0, \\[12pt]
    \mbox{At }Z=-1/2:\,\boldsymbol{U}=0 \mbox{ and } \Theta=1,\\[12pt]
    \mbox{At }Z=Z^*:\,\boldsymbol{W}=\boldsymbol{W}^*=W^*\,\vu{e}_z,\, \Theta_{p}=\Theta_{p}^*, \mbox{ and } \alpha=\mathcal{F}/\abs{\boldsymbol{W}^*},
    \end{array}\right\}
\end{equation}
where $Z^*$ is the vertical location of the particle injection. In the present work, we consider the injection of heavy ($\beta<1$) particles at the top cold surface $Z^*=1/2$, whereas the lighter ($\beta>1$) particles are injected at the bottom hot surface $Z^*=-1/2$. Here, $W^*$ is the particle injection velocity, and it can be negative (positive) for heavy (lighter) particles. The particle injection flux is represented by $\mathcal{F}$, which is always positive. The non-dimensional parameters are represented by
\begin{equation}
    \left.\begin{array}{l}
         \displaystyle Ra=\frac{g\,\beta_T\,\Delta T\,H^3}{\nu\,\kappa}, \quad Pr=\frac{\nu}{\kappa},\quad Ga=\frac{g\,H^3}{\nu^2}, \\[12pt]
         \displaystyle\Phi=\frac{d_p}{H},\quad\beta=\frac{3\,\rho}{\rho+2\,\rho_p},\quad E=\frac{\rho_p\,C_{p,\,p}}{\rho\,C_p},
    \end{array}
    \right\}
\end{equation}
where the parameters $Ra,\, Pr,\, Ga,\, \Phi,\, \beta$ and $E$ represent, respectively, the Rayleigh number, Prandtl number, Galilei number, particle size, density ratio, and specific heat capacity ratio between particle and fluid.

\section{Linear stability analysis}\label{sec:linear}

\subsection{Base state}
\begin{figure}
     \centering
     \begin{subfigure}[b]{0.49\textwidth}
         \centering
         \includegraphics[width=\textwidth]{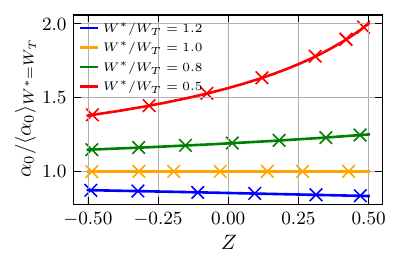}
         \caption{}
         \label{fig:heavy base-state}
     \end{subfigure}
     \hfill
    \begin{subfigure}[b]{0.49\textwidth}
         \centering
         \includegraphics[width=\textwidth]{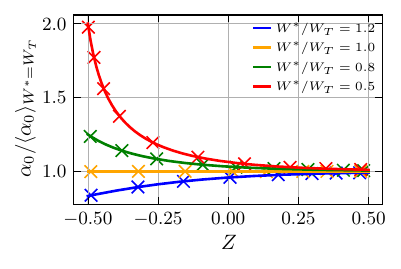}
         \caption{}
         \label{fig:light base-state}
     \end{subfigure}
    \captionsetup{width=\linewidth}
    \caption{\justifying Base-state particle volume fractions $\alpha_0(z)$ for (a) heavy particles ($\beta=0.5$) and (b) light particles ($\beta=3.0$) at various particle injection velocities. Here, the solid lines are the solution of (\ref{eq:W_0 solution}) and (\ref{eq:alpha_0 solution}), whereas the symbols are from~\cite{raza2025impact}. Other parameters kept at $Ga=1.92\times10^{10}$, $Pr=5$ and $\Phi=10^{-2}$.}
    \label{fig:base-state concentrations}
\end{figure}

Consider the following base state $\boldsymbol{U}=\boldsymbol{0},\, P=P_0(Z),\, \boldsymbol{W}=W_0(Z)\,\vu{e}_z,\, \alpha=\alpha_0(Z), \Theta=\Theta_0(Z)$ and $\Theta_p=\Theta_{p0}(Z)$.
\begin{align}\label{eq:base state alpha}
    &\mathcal{L}_\alpha\qty(\alpha_0,\, W_0)=D\qty(\alpha_0W_0)=0,\\
    &\mathcal{L}_z\qty(\alpha_0,\, P_0,\, W_0,\, \Theta_0,\, Ra,\, Pr,\, Ga,\, \beta,\, \Phi)\nonumber\\
    &\quad=DP_0-PrRa\,\Theta_0 -\frac{\alpha_0}{2}\qty[\qty(\beta-1)GaPr^2 + 12Pr\qty(3-\beta)\Phi^{-2}W_0]=0,\\ \label{eq:base state W}
    &\mathcal{L}_{pz}\qty(W_0,\, Ga,\, Pr,\, \beta,\, \Phi)=W_0DW_0+12Pr\beta\Phi^{-2}W_0+\qty(1-\beta)GaPr^2=0,\\ \label{eq:basestate theta equation}
    &\mathcal{L}_\theta\qty(\alpha_0,\, \Theta_0,\, \Theta_{p0},\, \Phi)=D^2\Theta_0-12\Phi^{-2}\alpha_0\qty(\Theta_0-\Theta_{p0})=0,\\ \label{eq:basestate thetap equation}
    &\mathcal{L}_{p\theta}\qty(W_0,\, \Theta_0,\, \Theta_{p0},\, E,\, \Phi)=W_0D\Theta_{p0}-12E^{-1}\Phi^{-2}\qty(\Theta_0-\Theta_{p0})=0
\end{align}
for $Z\in\qty(-1/2,\, 1/2)$. Where $D$ represents the derivative with respect to $Z$, and the boundary conditions are given by
\begin{equation}\label{eq:base state BCs}
    \left.\begin{array}{l}
    \mbox{At }Z=1/2:\,\Theta_0=0, \\[12pt]
    \mbox{At }Z=-1/2:\,\Theta_0=1,\\[12pt]
    \mbox{At }Z=Z^*:\,{W}_0=W^*,\, \Theta_{p0}=\Theta_{p}^*, \mbox{ and } \alpha_0=\mathcal{F}/\abs{W^*}.
    \end{array}\right\}
\end{equation}
It should be noted that (\ref{eq:base state alpha}) and (\ref{eq:base state W}) are nonlinear equations and can be solved analytically. Thus, the resulting solution after subjecting to boundary conditions is given by
\begin{gather}\label{eq:W_0 solution}
    W_0(Z) = W_T\qty[1+\mathscr{W}\qty((r-1)\exp{r-1}\exp{-(Z-Z^*)/l})],\\ \label{eq:alpha_0 solution}
    \alpha_0(Z)=\mathcal{F}/\abs{W_0},
\end{gather}
where $r=W^*/W_T$ with $W_T=-((1-\beta)/\beta)(\Phi^2/12)GaPr$ being the particle terminal speed, $l=\Phi^2\,W_T/\qty(12\,Pr\,\beta)$ and $\mathscr{W}(x)$ is the Lambert W function (which is the inverse of $f(x)=xe^x$). The resulting volume-fraction profiles are shown in figure~\ref{fig:base-state concentrations}. When the particles are injected at their terminal velocity ($W^*/W_T=1$) they keep that velocity across the layer and $\alpha_0$ is uniform. Particles injected more slowly accelerate towards $W_T$ and thin out as they do so, because the flux $\alpha_0W_0$ is conserved; particles injected faster decelerate and accumulate. The adjustment takes place over the relaxation length $l$. For the heavy particles of figure~\ref{fig:heavy base-state} this length is comparable to the layer depth, so $\alpha_0$ varies across the whole layer, whereas for the light particles of figure~\ref{fig:light base-state} the variation is confined to a region near the injection wall.

The basic state temperature profiles of fluid and particles are obtained numerically by solving the linear equations (\ref{eq:basestate theta equation}) and (\ref{eq:basestate thetap equation}) using $W_{0}$ and $\alpha_0$ from (\ref{eq:W_0 solution}) and (\ref{eq:alpha_0 solution}), respectively, and subjected to the boundary conditions (\ref{eq:base state BCs}).

\subsection{Linearized equations}
To examine the stability of the resulting basic state profile, we performed the classical normal mode analysis~\citep{drazin2004hydrodynamic} in which all the dependent variables are divided into their basic state plus their corresponding infinitesimal perturbations:
\begin{align}\label{eq:linear decomposition}
    \qty(P,\, \alpha,\, \boldsymbol{U}, \boldsymbol{W},\, \Theta,\, \Theta_p)=&\qty(P_0(Z),\, \alpha_0(Z),\, \boldsymbol{0},\, W_0(Z)\vu{e}_z,\, \Theta_0(Z),\, \Theta_{p0}(Z)) \nonumber\\
    &\quad\quad+ \qty(P',\, \alpha',\, \boldsymbol{U}',\, \boldsymbol{W}',\, \Theta',\, \Theta_p'),
\end{align}
where the variables with a prime superscript represent the infinitesimal perturbation quantities. By substituting (\ref{eq:linear decomposition}) in (\ref{eq:nondim fluid mass balance})--(\ref{eq:nondim particle energy balance}) and retaining only the linear terms, we obtain the following:

\begin{gather}\label{eq:perturbed fluid mass balance}
    \nabla\cdot\boldsymbol{U}'=0,\\
    \pdv{\alpha'}{\mathcal{T}}+W_0\pdv{\alpha'}{Z}+W_z'\,D\alpha_0=-\alpha_0\qty(\nabla\cdot\boldsymbol{W}')-\alpha'\,DW_0,\\ \nonumber
    \pdv{\boldsymbol{U}'}{\mathcal{T}}=-\nabla P'+Pr\nabla^2\boldsymbol{U}'+RaPr\,\Theta'\, \vu{e}_z\\ \nonumber
    +\frac{\alpha_0}{2}\qty[\qty(\beta-1)\pdv{\boldsymbol{U}'}{\mathcal{T}}-12Pr\,\qty(3-\beta)\Phi^{-2}\qty(\boldsymbol{U}'-\boldsymbol{W}')] \nonumber \\
    +\frac{\alpha'}{2}\qty(\beta-1)GaPr^2\vu{e}_z+\frac{6Pr\qty(3-\beta)W_0}{\Phi^2}\,\alpha'\,\vu{e}_z,\\
     \pdv{\boldsymbol{W}'}{\mathcal{T}}+W_0\pdv{\boldsymbol{W}'}{Z}+\qty(DW_0)\,{W}'_z\vu{e}_z=\beta \pdv{\boldsymbol{U}'}{\mathcal{T}} + {12Pr\beta}\,{\Phi^{-2}}\qty(\boldsymbol{U}'-\boldsymbol{W}'),\\
    \pdv{\Theta'}{\mathcal{T}}+\qty(D\Theta_0)U_z'=\nabla^2\Theta'+\alpha_0\qty[\pdv{\Theta'}{\mathcal{T}}+\qty(D\Theta_0)U_z'-12\Phi^{-2}\qty({\Theta'-\Theta_p'})]\nonumber\\
    -12\Phi^{-2}\qty(\Theta_0-\Theta_{p0})\alpha',\\ \label{eq:perturbed particle energy balance}
    \pdv{\Theta_p'}{\mathcal{T}}+W_0\,\pdv{\Theta_p'}{Z}+\qty(D\Theta_{p0})\, W_z'=12E^{-1}\Phi^{-2}\qty(\Theta'-\Theta_{p}').
\end{gather}
As the above set of differential equations is linear with coefficients that depend only on $\boldsymbol{x}$, not $t$, they admit a general solution in the form of a normal mode:
\begin{equation}\label{eq:normal mode form}
    \qty(P',\, \alpha',\, \boldsymbol{U}',\, \boldsymbol{W}',\, \Theta',\, \Theta_p')^T = e^{\text{i}\mathcal{K}\qty(X-\mathcal{C}\mathcal{T})} \qty(\widehat{P}(Z),\, \widehat{\alpha}(Z),\, \widehat{\boldsymbol{U}}(Z),\, \widehat{\boldsymbol{W}}(Z),\, \widehat{\Theta}(Z),\, \widehat{\Theta}_p(Z))^T,
\end{equation}
where hat quantities are vertical structure functions of the respective variables, $\mathcal{K}$ is the wavenumber, and $\mathcal{C}=\mathcal{C}_r+\text{i}\,\mathcal{C}_i$ is the complex wave speed. The imaginary part $\mathcal{C}_i$ represents an instability with exponential growth when $\mathcal{C}_i>0$ or stability with an exponential decay when $\mathcal{C}_i<0$ or neutral/marginal stability when $\mathcal{C}_i=0$. Substitution of (\ref{eq:normal mode form}) in (\ref{eq:perturbed fluid mass balance})--(\ref{eq:perturbed particle energy balance}) leads to a system of linear perturbation equations in operator form, as listed in \S~\ref{appen:linear perturbed system}. The boundary conditions are given by
\begin{equation}\label{eq:linear BCs}
    \left.\begin{array}{l}
    \mbox{At }Z=1/2:\,\widehat{U}_x=\widehat{U}_z=0 \mbox{ and } \widehat{\Theta}=0, \\[12pt]
    \mbox{At }Z=-1/2:\,\widehat{U}_x=\widehat{U}_z=0 \mbox{ and } \widehat{\Theta}=0,\\[12pt]
    \mbox{At }Z=Z^*:\,\widehat{W}_x=\widehat{W}_z=0,\, \widehat{\Theta}_{p}=0, \mbox{ and } \widehat{\alpha}=0.
    \end{array}\right\}
\end{equation}
The system of equations (\ref{eq:linear fluid mass balance})--(\ref{eq:linear particle energy balance}) together with the above set of boundary conditions gives rise to the following generalised eigenvalue problem,
\begin{equation}
    \mathscr{A}\boldsymbol{Q}=\mathcal{C}\mathscr{B}\boldsymbol{Q},
\end{equation}
where $\mathscr{A}$ and $\mathscr{B}$ are complex square matrices, $\mathcal{C}$ is the eigenvalue corresponding to the eigenfunction $\boldsymbol{Q}=\qty(\widehat{P},\, \widehat{\alpha},\, \widehat{U}_x,\, \widehat{U}_z,\, \widehat{W}_x,\, \widehat{W}_z,\, \widehat{\Theta},\, \widehat{\Theta}_p)$.

\subsection{Review of linear stability results}
To elucidate the various mechanisms that contribute to stabilization/destabilization of the system, we derived a quadratic energy balance equation for the linear perturbation field $\widehat{U}_z(z)$ at the neutral stability condition as
\begin{equation}\label{eq:normalised linera energy balance}
    E^\Theta+E^{\alpha_0W}+E^{\alpha W_0}+E^{\alpha Ga}=1,
\end{equation}
where $E^{\Theta}, E^{\alpha W_0}$, $E^{\alpha_0 W}$ and $E^{\alpha Ga}$ represent the relative strength of the power of the thermal buoyancy, the rate of work done by the particle due to non-uniform base-state setting velocity, the rate of work done by the particle due to non-uniform base-state volume fraction, and the rate of work done by Archimedes’ force, respectively. Here, the mathematical expressions for $E^{\Theta}$, $E^{\alpha_0W}$, $E^{\alpha W_0}$ and $E^{\alpha Ga}$ are given by
\refstepcounter{equation}
$$
 E^{\Theta}=\displaystyle\frac{e^{\Theta}}{\abs{e^{V}}},\quad E^{\alpha_0W}=\displaystyle\frac{e^{\alpha_0 W}}{\abs{e^{V}}},\quad E^{\alpha W_0}=\displaystyle\frac{e^{\alpha W_0}}{\abs{e^{V}}},\quad E^{\alpha Ga}=\displaystyle\frac{e^{\alpha Ga}}{\abs{e^{V}}},
\eqno{(\theequation{\mathit{a},\,\mathit{b},\, \mathit{c},\, \mathit{d}})}
$$
where $e^{V}$ represents the rate of viscous dissipation, and other terms $e^{\Theta},\, e^{\alpha_0W},\, e^{\alpha W_0},\, e^{\alpha Ga}$ are given along with their physical interpretations in~\S\ref{appen:linear KE equations}.

\begin{figure}
     \centering
     \begin{subfigure}[b]{0.49\textwidth}
         \centering
         \includegraphics[width=\textwidth]{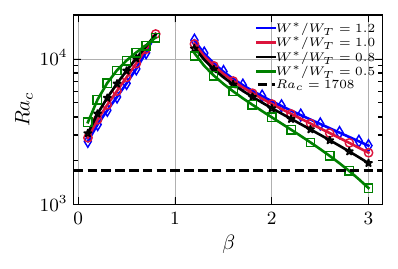}
         \caption{}
         \label{fig:beta vs Rac}
     \end{subfigure}
     \hfill
    \begin{subfigure}[b]{0.49\textwidth}
         \centering
         \includegraphics[width=\textwidth]{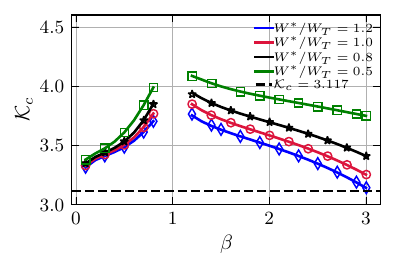}
         \caption{}
         \label{fig:beta vs Kc}
     \end{subfigure}
     \captionsetup{width=\linewidth}
    \caption{\justifying Variation of critical Rayleigh number $Ra_c$ (a) and critical wavenumber $\mathcal{K}_c$ (b) with the density ratio $\beta$ at various particle injection velocities. Here, the solid lines are from the present study, whereas the symbols are from~\cite{raza2025impact}. Other parameters kept at $Ga=1.92\times10^{10}$, $Pr=5$, $\Phi=10^{-2}$, particle injection flux $\mathcal{F}=\mathcal{F}_0=533.3$ and heat capacity ratio $E=5\times 10^{-3}$. The horizontal dashed lines in (a) and (b) represent the single-phase $Ra_c$ and $\mathcal{K}_c$, respectively.}
    \label{fig:beta vs Rac and Kc}
\end{figure}

\begin{figure}
     \centering
     \begin{subfigure}[b]{0.49\textwidth}
         \centering
         \includegraphics[width=\textwidth]{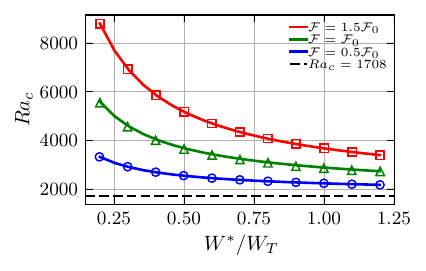}
         \caption{}
         \label{fig:Wstar vs Rac beta 0.1}
     \end{subfigure}
     \hfill
    \begin{subfigure}[b]{0.49\textwidth}
         \centering
         \includegraphics[width=\textwidth]{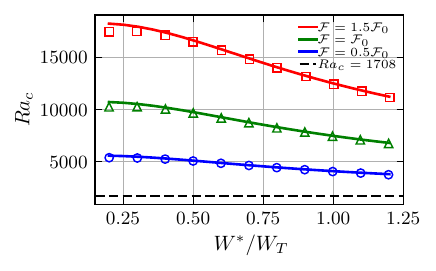}
         \caption{}
         \label{fig:Wstar vs Rac beta 0.5}
     \end{subfigure}
     \hfill
    \begin{subfigure}[b]{0.49\textwidth}
         \centering
         \includegraphics[width=\textwidth]{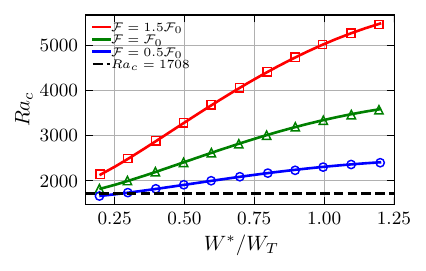}
         \caption{}
         \label{fig:Wstar vs Rac beta 2.5}
     \end{subfigure}
     \hfill
    \begin{subfigure}[b]{0.49\textwidth}
         \centering
         \includegraphics[width=\textwidth]{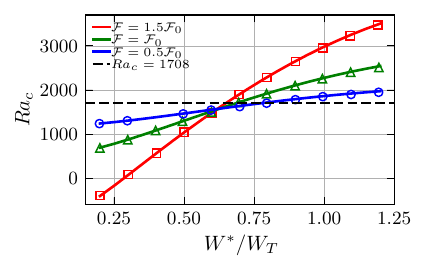}
         \caption{}
         \label{fig:Wstar vs Rac beta 3.0}
     \end{subfigure}
     \captionsetup{width=\linewidth}
    \caption{\justifying Variation of critical Rayleigh number $Ra_c$ with the particle inlet velocity $W^*/W_T$ at various density ratios: (a) $\beta=0.1$, (b) $\beta=0.5$, (c) $\beta=2.5$ and (d) $\beta=3.0$. Here, the solid lines are from the present study, whereas the symbols are from~\cite{raza2025impact}. Other parameters kept at $Ga=1.92\times10^{10}$, $Pr=5$, $\Phi=10^{-2}$, particle injection flux $\mathcal{F}=\mathcal{F}_0=533.3$ and heat capacity ratio $E=5\times 10^{-3}$. The horizontal dashed line in all graphs represents the single-phase $Ra_c$.}
    \label{fig:Wstar vs Rac}
\end{figure}

\begin{figure}
     \centering
     \begin{subfigure}[b]{0.49\textwidth}
         \centering
         \includegraphics[width=\textwidth]{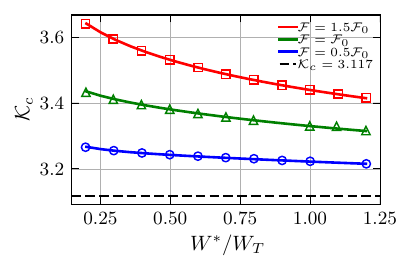}
         \caption{}
         \label{fig:Wstar vs Kc beta 0.1}
     \end{subfigure}
     \hfill
    \begin{subfigure}[b]{0.49\textwidth}
         \centering
         \includegraphics[width=\textwidth]{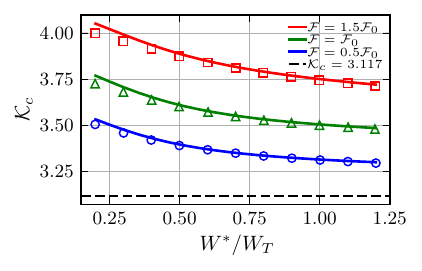}
         \caption{}
         \label{fig:Wstar vs Kc beta 0.5}
     \end{subfigure}
     \hfill
    \begin{subfigure}[b]{0.49\textwidth}
         \centering
         \includegraphics[width=\textwidth]{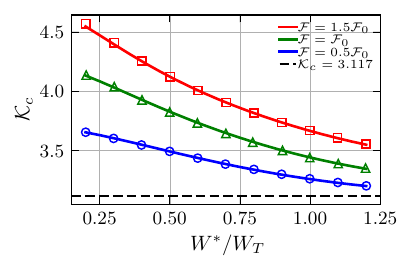}
         \caption{}
         \label{fig:Wstar vs Kc beta 2.5}
     \end{subfigure}
     \hfill
    \begin{subfigure}[b]{0.49\textwidth}
         \centering
         \includegraphics[width=\textwidth]{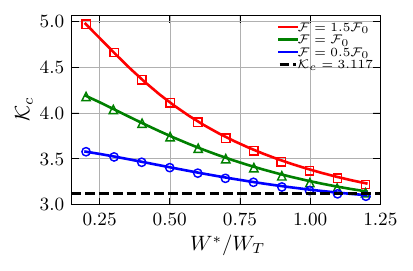}
         \caption{}
         \label{fig:Wstar vs Kc beta 3.0}
     \end{subfigure}
     \captionsetup{width=\linewidth}
    \caption{\justifying Variation of critical wavenumber $\mathcal{K}_c$ with the particle inlet velocity $W^*/W_T$ at various density ratios: (a) $\beta=0.1$, (b) $\beta=0.5$, (c) $\beta=2.5$ and (d) $\beta=3.0$. Here, the solid lines are from the present study, whereas the symbols are from~\cite{raza2025impact}. Other parameters kept at $Ga=1.92\times10^{10}$, $Pr=5$, $\Phi=10^{-2}$, particle injection flux $\mathcal{F}=\mathcal{F}_0=533.3$ and heat capacity ratio $E=5\times 10^{-3}$. The horizontal dashed line in all graphs represents the single-phase $\mathcal{K}_c$.}
    \label{fig:Wstar vs Kc}
\end{figure}

\begin{figure}
     \centering
     \begin{subfigure}[b]{0.49\textwidth}
         \centering
         \includegraphics[width=\textwidth]{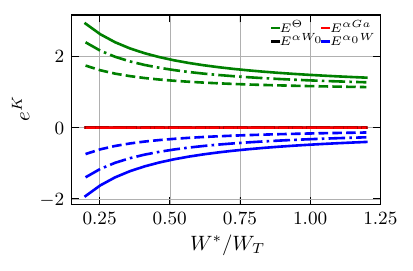}
         \caption{}
         \label{fig:Wstar vs energy beta 0.1}
     \end{subfigure}
     \hfill
    \begin{subfigure}[b]{0.49\textwidth}
         \centering
         \includegraphics[width=\textwidth]{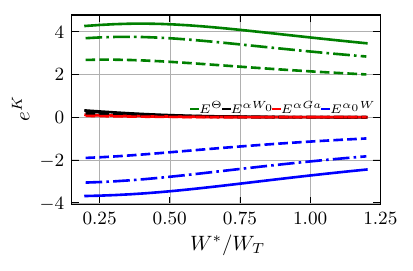}
         \caption{}
         \label{fig:Wstar vs energy beta 0.5}
     \end{subfigure}
     \hfill
    \begin{subfigure}[b]{0.49\textwidth}
         \centering
         \includegraphics[width=\textwidth]{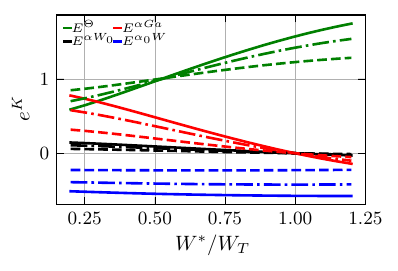}
         \caption{}
         \label{fig:Wstar vs energy beta 2.5}
     \end{subfigure}
     \hfill
    \begin{subfigure}[b]{0.49\textwidth}
         \centering
         \includegraphics[width=\textwidth]{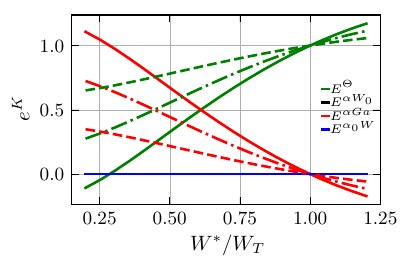}
         \caption{}
         \label{fig:Wstar vs energy beta 3.0}
     \end{subfigure}
     \captionsetup{width=\linewidth}
    \caption{\justifying Variation of each term in~\eqref{eq:normalised linera energy balance} with the particle inlet velocity $W^*/W_T$ at various density ratios: (a) $\beta=0.1$, (b) $\beta=0.5$, (c) $\beta=2.5$ and (d) $\beta=3.0$. Here, dashed, dashed-dotted, and solid lines correspond to $\mathcal{F}/\mathcal{F}_0=0.5,\, 1.0,$ and $1.5$, respectively. Other parameters kept at $Ga=1.92\times10^{10}$, $Pr=5$, $\Phi=10^{-2}$, and $E=5\times 10^{-3}$.}
    \label{fig:Wstar vs energy}
\end{figure}

The linear results are summarised in figures~\ref{fig:beta vs Rac and Kc}--\ref{fig:Wstar vs energy}, where the solid lines are our calculations and the symbols are taken from \citet{raza2025impact}. The two sets agree closely, which gives us confidence in the linear operator and its eigenfunctions that the weakly nonlinear analysis is built on. Unless stated otherwise we use $Ga=1.92\times10^{10}$, $Pr=5$, $\Phi=10^{-2}$ and $E=5\times10^{-3}$, and the particles are released at the temperature of the injection wall. For every density ratio in figure~\ref{fig:beta vs Rac} the critical Rayleigh number lies above the single-phase value, and the stabilisation is strongest for nearly neutrally buoyant particles. This last point is a consequence of fixing the flux rather than the concentration: as $\beta\to1$ the terminal velocity vanishes, so the volume fraction $\mathcal{F}/\abs{W_T}$ needed to carry a given flux grows without bound. The critical wavenumber in figure~\ref{fig:beta vs Kc} follows the same trend and is always larger than the single-phase value $\mathcal{K}_c=3.117$.

The effect of the injection velocity is shown in figures~\ref{fig:Wstar vs Rac} and \ref{fig:Wstar vs Kc}. For heavy particles, slow injection raises $Ra_c$, and more so at larger flux. For light particles, the trend is reversed: slow injection lowers $Ra_c$, and for $\beta=3$ at $\mathcal{F}=1.5\mathcal{F}_0$ the critical Rayleigh number falls below the single-phase value and becomes negative for $W^*/W_T\lesssim0.3$ (figure~\ref{fig:Wstar vs Rac beta 3.0}). The energy budget in figure~\ref{fig:Wstar vs energy} explains this asymmetry. For heavy particles, the buoyancy term $E^\Theta$ exceeds unity and is balanced by a negative $E^{\alpha_0W}$, the work done by the drag in a layer of non-uniform particle concentration; the Archimedes term $E^{\alpha Ga}$ and the settling-velocity term $E^{\alpha W_0}$ are negligible. For light particles, the Archimedes term is the one that matters. It is positive, and hence destabilising, when the particles are injected below their terminal speed; it vanishes at $W^*=W_T$ where the base state is uniform, and it changes sign beyond. At $\beta=3$ and high flux, it becomes large enough that $E^\Theta$ itself turns negative: the convective mode is then driven by the particles rather than by the thermal buoyancy, which is why the layer can become unstable when heated from above.

\section{Finite amplitude formulation}\label{sec:finite amplitude}
Following the previous works on finite amplitude analysis by~\cite{stuart1960non, yao1992finite, khandelwal2015weakly, sharma2018finite, aleria2024finite, srinivas2025weakly}, we expand all the dependent variables in separable form with the most unstable growth rate $\mathcal{C}_i$ as a small parameter. For example, the fluid temperature field $\Theta$ is written as
\begin{equation}\label{eq:Theta perturbaton series}
    \begin{split}
        \Theta(X,\, Z,\, \mathcal{T})&=\mathbb{E}^0\qty[\Theta_0(Z)+\mathcal{C}_i\abs{B(\tau)}^2\Theta_1(Z)+\order{\mathcal{C}_i^2}]\\
        &\quad+\mathbb{E}^1\qty[\mathcal{C}_i^{1/2}B(\tau)\Theta_{10}(Z)+\mathcal{C}_i^{3/2}B(\tau)\abs{B(\tau)}^2\Theta_{11}(Z)+\order{\mathcal{C}_i^{5/2}}]\\
        &\quad\quad+\mathbb{E}^2\qty[\mathcal{C}_iB(\tau)^2\Theta_{20}(Z)+\order{\mathcal{C}_i^2}]+\dots+\text{c.c.},
    \end{split}
\end{equation}
where the harmonic function $\mathbb{E}^j=\exp(j\qty[\text{i}\mathcal{K}_c\qty(X-\mathcal{C}_r\mathcal{T})])$ for $j=0,\, 1,\ 2$, the wavenumber and wave speed corresponding to the critical Rayleigh number $Ra_c$ are represented by $\mathcal{K}_c$ and $\mathcal{C}_r$, respectively. Here, c.c. represents the complex conjugate of only complex-valued functions, and $B(\tau)$ is the unknown amplitude function, which is assumed to vary slowly at a time scale $\tau=\mathcal{C}_i\mathcal{T}$. As shown in previous work~\citep{srinivas2025weakly}, we decompose the temporal variation of each dependent variable into two parts: one is fast variation at a time scale $\mathcal{T}$ and the second is the slow variation at time scale $\tau$. That is, a function $f(\mathcal{T})$ can be written as $f(\mathcal{T})=F(\mathcal{T},\, \tau)$, so that its time derivative becomes
\begin{equation}
    \pdv{f}{\mathcal{T}}=\pdv{F}{\mathcal{T}} + \mathcal{C}_i\pdv{F}{\tau},
\end{equation}
where $\mathcal{T}$ and $\tau$ in $F(\tau,\, \mathcal{T})$ are treated as independent variables. This splitting of time scales leads to a method of multiple time scales, which gives an evolution equation for the unknown amplitude function called the Landau cubic equation. For justification of the expansion given in~\eqref{eq:Theta perturbaton series}, we refer to the previous work by~\cite{srinivas2025weakly}.

\subsection{Derivation of cubic Landau equation}
We note that the first Landau coefficient does not depend on the higher harmonics such as $\mathbb{E}^3,\, \mathbb{E}^4,\, \dots$, and they are neglected in the expansion~\eqref{eq:Theta perturbaton series}. By substituting similar expansions of the form~\eqref{eq:Theta perturbaton series} for all dependent variables in \eqref{eq:nondim fluid mass balance}--\eqref{eq:nondim particle energy balance}, we obtain the equations for different harmonic components, which are listed in \S~\ref{appen:harmonic equations}. The equations listed in \S~\ref{appen:harmonic equations} can be solved in increasing order of growth rate $\mathcal{C}_i$. First, the equations at $\order{\mathcal{C}_i^0}$ \eqref{eq:E0 alpha}--\eqref{eq:E0 thetap} are solved, which are similar to the base state equations \eqref{eq:base state alpha}--\eqref{eq:basestate thetap equation}. Second, the equations at $\order{\mathcal{C}_i^{1/2}}$ \eqref{eq:E1 fluid mass balance}--\eqref{eq:E1 thetap} are solved for the eigenfunctions $P_{10},\, \alpha_{10},\, U_{x10},\, U_{z10},\, W_{x10},\, W_{z10},\, \Theta_{10},\, \Theta_{p10}$, which are similar to the linear stability perturbation equations given by~\eqref{eq:linear fluid mass balance}--\eqref{eq:linear particle energy balance} at the critical wavenumber $\mathcal{K}_c$ and Rayleigh number $Ra_c$. Third, the inhomogeneous equations at $\order{\mathcal{C}_i}$ are solved, which have contributions from the harmonics $\mathbb{E}^0$ and $\mathbb{E}^2$. At this order, as shown in~\eqref{eq:E0 alpha}--\eqref{eq:E0 thetap} and \eqref{eq:E2 mass balance}--\eqref{eq:E2 thetap}, both harmonics contain the known inhomogeneous terms that arise due to the nonlinear interaction of the fundamental harmonics $E^1$ at $\order{\mathcal{C}_i^{1/2}}$. We call the equations corresponding to the zeroth harmonic the distortion of the base state, or the Reynolds stress terms $\alpha_1,\, W_{z1},\, \Theta_1,\, \Theta_{p1}$. Finally, the singular inhomogeneous equations at $\order{\mathcal{C}_i^{3/2}}$ are solved as follows: the system of equations at this order shares the same linear stability operator defined at $\order{\mathcal{C}_i^{1/2}}$ of harmonic $\mathbb{E}^1$, which satisfies homogeneous equations. However, at $\order{\mathcal{C}_i^{3/2}}$, the system is inhomogeneous with unknown terms $dB/d\tau$, $B$, $B\abs{B}^2$ and the known terms from the lower orders. Hence, to obtain a non-trivial solution for the eigenfunctions $P_{11},\, \alpha_{11},\, U_{x11},\, U_{z11},\, W_{x11},\, W_{z11},\, \Theta_{11},\, \Theta_{p11}$, we impose a Fredholm solvability condition to the
right-hand inhomogeneous terms. This requires the solution to the system of adjoint equations (see~\S \ref{appen:adjoint linear system}) corresponding to the linear stability operator. Accordingly, the inhomogeneous part should be orthogonal to the adjoint eigenfunctions $P^\dag,\, \alpha^\dag,\, U_{x}^\dag,\, U_z^\dag,\, W_x^\dag,\, W_z^\dag,\, \Theta^\dag,\, \Theta_p^\dag$, which leads to the following Landau cubic equation
\begin{equation}\label{eq:Landau equation}
    \dv{B}{\tau}=\mathcal{K}_cB+a_1B\abs{B}^2,
\end{equation}
where $a_1$ is the first Landau constant defined as
\begin{equation}\label{eq:a1}
    a_1=-\int_{Z=-1/2}^{1/2}\qty{\mathcal{G}_\alpha\alpha^\dag+\mathcal{G}_xU_x^\dag+\mathcal{G}_zU_z^\dag+\mathcal{G}_{px}W_x^\dag+\mathcal{G}_{pz}W_z^\dag+\mathcal{G}_{\theta}\Theta^\dag+\mathcal{G}_{p\theta}\Theta_p^\dag}\,dZ,
\end{equation}
in the above, the scalar functions $\qty(\mathcal{G}_\alpha,\, \mathcal{G}_x,\, \mathcal{G}_z,\, \mathcal{G}_{px},\, \mathcal{G}_{pz},\, \mathcal{G}_{\theta},\, \mathcal{G}_{p\theta})$ are given by~(\ref{eq:G_alpha})--(\ref{eq:G_ptheta}) in \S~\ref{appen:scalar functions}. By changing the variables $A=\mathcal{C}_i^{1/2}B$ and $\tau=\mathcal{C}_i\mathcal{T}$, and straight forward manipulations, \eqref{eq:Landau equation} can be written as
\begin{equation}\label{eq:A^2 ode}
    \dv{\abs{A}^2}{\mathcal{T}}=2\mathcal{K}_c\mathcal{C}_i\abs{A}^2+2\Re{a_1}\abs{A}^4,
\end{equation}
where $\Re{a_1}$ is the real of $a_1$. The above equation has the following three stationary points
\refstepcounter{equation}
$$
    A_e=0, \quad\mbox{and}\quad A_e=\pm\sqrt{-\mathcal{K}_c\mathcal{C}_i/\Re{a_1}}.\label{eq:equilibrium conditions}
     \eqno{(\theequation{\mathit{a},\mathit{b}})}
$$
The stationary point ${A_e}=0$ corresponds to the steady base state solution, which is linearly stable when $Ra<Ra_c$, with $Ra_c$ as a critical Rayleigh number obtained from the linear stability analysis. Hence, from~(\ref{eq:equilibrium conditions}$\mathit{b}$) it is clear that the finite amplitude solution exists only if $\Re{a_1}$ and $\mathcal{C}_i$ have opposite signs~\citep{drazin2004hydrodynamic, shukla2011weakly}. This is possible in two distinct ways: first with $\Re{a_1}<0$ and $\mathcal{C}_i>0$ (for $Ra>Ra_c$); second with  $\Re{a_1}>0$ and $\mathcal{C}_i<0$ (for $Ra<Ra_c$). The first possibility leads to a supercritical pitchfork bifurcation (from here on wards, supercritical bifurcation) and the second one leads to a subcritical pitchfork bifurcation (from here on wards, subcritical bifurcation). The stationary solution of~\eqref{eq:A^2 ode} corresponding to the supercritical bifurcation is called the equilibrium amplitude, whereas the stationary solution of~\eqref{eq:A^2 ode} corresponding to the subcritical bifurcation is called the threshold amplitude. By using the equilibrium amplitude from (\ref{eq:equilibrium conditions}$\mathit{b}$), the solution to~\eqref{eq:A^2 ode} for supercritical bifurcation is given by
\begin{equation}\label{eq:A^2 solution}
    \abs{A}^2=\frac{A_e^2}{1+\displaystyle\qty(\frac{\abs{A_e}^2}{\abs{A_0}^2}-1)e^{-2\mathcal{K}_c\mathcal{C}_i\mathcal{T}}},
\end{equation}
where $\abs{A_0}$ is $\abs{A}$ at $\mathcal{T}=0$. Hence, from~\eqref{eq:A^2 solution}, it is evident that as $\mathcal{T}\to\infty$, $\abs{A}\to A_e$ for all finite $\abs{A_0}$. Using the expression for $\abs{A}$, the closed form solution for $A(\mathcal{T})$~\citep{srinivas2025weakly} is given by
\begin{equation}
    \frac{A}{A_0}=\frac{\abs{A}}{\abs{A_0}}\qty{\frac{\abs{A}}{\abs{A_0}}e^{-\mathcal{K}_c\mathcal{C}_i\mathcal{T}}}^{\mbox{i}\Im{a_1}/\Re{a_1}},
\end{equation}
where $A_0$ is the initial value of $A$ and $\Im{a_1}$ is the imaginary part of $a_1$.

\section{Results and discussion}\label{sec:results}
We now use the Landau coefficient to classify the bifurcation over the range of parameters covered by the linear analysis. The fluid and particle properties are those of \S~\ref{sec:linear}: $Ga=1.92\times10^{10}$, $Pr=5$, $\Phi=10^{-2}$ and $E=5\times10^{-3}$, with heavy particles released at the temperature of the cold upper wall and light particles at that of the hot lower wall. The flux is varied around the reference value $\mathcal{F}_0=533.3$ used by \citet{raza2025impact}, and the injection velocity is varied between $0.2W_T$ and $1.2W_T$. Distance from onset is measured by the reduced Rayleigh number
\begin{equation}\label{eq:reduced Ra}
\delta_{Ra}=\begin{cases}
     \qty(Ra-Ra_c)/Ra_c, & \quad\mbox{for}\quad Ra_c\geq 0,\\
      \qty(Ra_c-Ra)/Ra_c, & \quad\mbox{for}\quad Ra_c<0.
\end{cases}
\end{equation}
The second form is needed for the light-particle cases in which $Ra_c$ is negative (see figure~\ref{fig:Wstar vs Rac beta 3.0} for $\mathcal{F}=1.5\mathcal{F}_0$); with this convention $\delta_{Ra}>0$ always lies on the linearly unstable side of the threshold. All the Landau coefficients reported below are evaluated at $\delta_{Ra}=5\times10^{-3}$.

\subsection{Numerical method and convergence}\label{sec:numerics}
All the problems of \S~\ref{sec:finite amplitude}, namely the linear eigenvalue problem, the mean-flow distortion, the second harmonic and the adjoint problem, are discretised by Chebyshev collocation on $N$ Gauss--Lobatto points using the differentiation matrices of \citet{weideman2000matlab}. The integrals in \eqref{eq:a1} are evaluated by Gauss--Legendre quadrature after barycentric interpolation of the integrands. At the critical point, the leading eigenvalue is purely imaginary, $\mathcal{C}_r=0$, so the onset is stationary, as reported by \citet{raza2024stabilization} for the same model. The fundamental eigenfunction is normalised so that $\int_{-1/2}^{1/2}\sum_j\abs{q_{j,10}}^2\,\mathrm{d}Z=1$, where the sum runs over $\alpha_{10}$, $U_{x10}$, $U_{z10}$, $W_{x10}$, $W_{z10}$, $\Theta_{10}$ and $\Theta_{p10}$, and the adjoint eigenfunction is then fixed by \eqref{eq:adjoint normalisation}. The equilibrium amplitude $A_e$ depends on the first of these choices, while the sign of $\Re{a_1}$, which is what decides the type of bifurcation, does not.

Table~\ref{tab:grid independence test} shows how the growth rate, the real part of the Landau coefficient, and the equilibrium amplitude change with $N$ for $\beta=0.5$ and three injection velocities. The values agree to about eight significant figures between $N=30$ and $N=60$, and the remaining scatter is at the level of round-off in the generalised eigenvalue solver rather than a systematic trend. All results below were obtained with $N=45$. The table also shows a first trend that we return to below: for this heavy-particle case, injecting the particles faster makes $\Re{a_1}$ more negative and therefore lowers the saturated amplitude of the convection.

To separate the physical origin of the different contributions, we split $a_1$ according to the balance equation in which each nonlinear term originates,
\begin{align}\label{eq:a_1 split}
a_1 = a_{1\alpha} + a_{1X} + a_{1Z} + a_{1X_p} + a_{1Z_p} + a_{1\Theta} + a_{1\Theta_p},
\end{align}
where
\begin{equation}\label{eq:a_1 terms}
    \left.\begin{array}{l}
a_{1\alpha} = -\int_{-1/2}^{1/2} \mathcal{G}_\alpha \alpha^\dagger \,\mathrm{d}Z,\;
a_{1X} = -\int_{-1/2}^{1/2} \mathcal{G}_x U_x^\dagger \,\mathrm{d}Z,\;
a_{1Z} = -\int_{-1/2}^{1/2} \mathcal{G}_z U_z^\dagger \,\mathrm{d}Z,\\ [12pt]
a_{1X_p} = -\int_{-1/2}^{1/2} \mathcal{G}_{px} W_{x}^\dagger \,\mathrm{d}Z,\;
a_{1Z_p} = -\int_{-1/2}^{1/2} \mathcal{G}_{pz} W_{z}^\dagger \,\mathrm{d}Z,\;
a_{1\Theta} = -\int_{-1/2}^{1/2} \mathcal{G}_{\theta} {\Theta}^\dagger \,\mathrm{d}Z,\\ [12pt]
\quad\quad\quad\quad\quad\quad\quad\quad\quad\quad\quad a_{1\Theta_p} = -\int_{-1/2}^{1/2} \mathcal{G}_{p\theta}{\Theta_p}^\dagger \,\mathrm{d}Z.
    \end{array}\right\}
\end{equation}
The first three terms come from the fluid mass and momentum equations, $a_{1X_p}$ and $a_{1Z_p}$ from the particle momentum equation, $a_{1\alpha}$ from the particle continuity equation, and $a_{1\Theta}$ and $a_{1\Theta_p}$ from the fluid and particle energy equations. Each term depends on the base state, the fundamental mode, the mean distortion and the second harmonic, so this is not a clean separation of physical mechanisms. It does, however, tell us which balance equation the sign of $a_1$ is decided in.

\begin{table}
\centering
\begin{tabular}{ccccc}
$W^*/W_T$                     & $N$  & $\mathcal{C}_i$ & $\Re{a_1}$ & $A_e$ \\
\midrule
\multirow{4}{*}{$0.5$} & 30 & 0.069367581400273978 & -0.24824079151029294  & 1.0042915753697088     \\
                             & 40 & 0.069367582681731113 & -0.24824078128888777  & 1.0042916053221005     \\
                             & 50 & 0.069367575693201428 & -0.24824079391524739  & 1.0042915291919310     \\
                             & 60 & 0.069367584740223961 & -0.24824078891127799  & 1.0042916048046440     \\ [12pt]
\midrule
\multirow{4}{*}{$0.8$} & 30 & 0.069084210982896840  & -0.25820185436998855  & 0.97219697999144072    \\
                             & 40 & 0.069084208348888829  & -0.25820184787789058  & 0.97219697367993874    \\
                             & 50 & 0.069084205101482138  & -0.25820186539529255  & 0.97219691785137097    \\
                             & 60 & 0.069084214582689846  & -0.25820185006126301  & 0.97219701343246456    \\ [12pt]
\midrule
\multirow{4}{*}{$1.2$} & 30 & 0.068821691942773780  & -0.28074055236726680  & 0.92427325204816202    \\
                             & 40 & 0.068821678079622042  & -0.28074049578591331  & 0.92427325209780808    \\
                             & 50 & 0.068821676408455559  & -0.28074055651594176  & 0.92427314090616808    \\
                             & 60 & 0.068821682984059412  & -0.28074053277674560  & 0.92427322413915458   \\
\bottomrule
\end{tabular}
\captionsetup{width=\linewidth}
\caption{\justifying Grid independence test on growth rate $(\mathcal{C}_i)$, real part of Landau constant $\left(\Re{a_1}\right)$ and equilibrium amplitude $(A_e)$ for $Ga=1.92\times10^{10}$, $Pr=5$, $E=5\times10^{-3}$, $\Phi=10^{-2}$, $\beta=0.5$, $\mathcal{F}=\mathcal{F}_0$ and $W^*/W_T=\qty{0.5,\, 0.8,\, 1.2}$ at $\delta_{Ra}=0.01$.}
\label{tab:grid independence test}
\end{table}

\subsection{Effect of the injection velocity}\label{sec:injection velocity}
\begin{figure}
     \centering
     \begin{subfigure}[b]{0.49\textwidth}
         \centering
         \includegraphics[width=\textwidth]{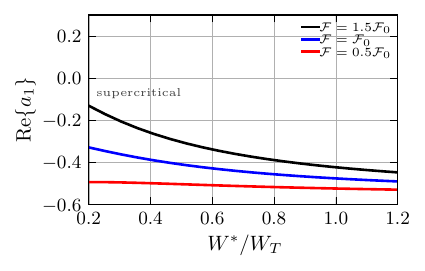}
         \caption{}
         \label{fig:Wstar vs a1r beta 0.1}
     \end{subfigure}
     \hfill
    \begin{subfigure}[b]{0.49\textwidth}
         \centering
         \includegraphics[width=\textwidth]{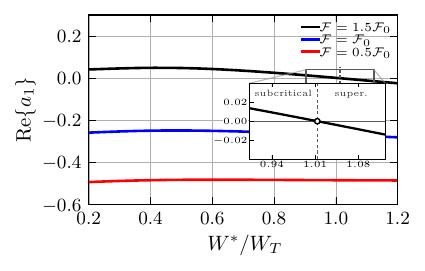}
         \caption{}
         \label{fig:Wstar vs a1r beta 0.5}
     \end{subfigure}
     \hfill
    \begin{subfigure}[b]{0.49\textwidth}
         \centering
         \includegraphics[width=\textwidth]{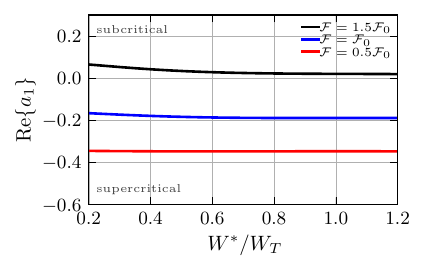}
         \caption{}
         \label{fig:Wstar vs a1r beta 2.5}
     \end{subfigure}
     \hfill
    \begin{subfigure}[b]{0.49\textwidth}
         \centering
         \includegraphics[width=\textwidth]{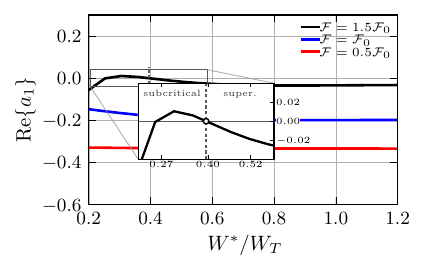}
         \caption{}
         \label{fig:Wstar vs a1r beta 3.0}
     \end{subfigure}
     \captionsetup{width=\linewidth}
    \caption{\justifying Variation in real part of Landau constant given by~\eqref{eq:a1} with the particle inlet velocity $W^*/W_T$ at various density ratios: (a) $\beta=0.1$, (b) $\beta=0.5$, (c) $\beta=2.5$ and (d) $\beta=3.0$. Other parameters kept at $Ga=1.92\times10^{10}$, $Pr=5$, $\Phi=10^{-2}$, $E=5\times 10^{-3}$ and $\delta_{Ra}=5\times10^{-3}$. The insets in (b) and (d) magnify the region where $\Re{a_1}$ changes sign for $\mathcal{F}=1.5\mathcal{F}_0$.}
    \label{fig:Wstar vs a1r}
\end{figure}

Figure~\ref{fig:Wstar vs a1r} shows $\Re{a_1}$ as a function of the injection velocity for two heavy and two light particle types and three values of the flux. At the two lower fluxes, the coefficient is negative everywhere, so the bifurcation is supercritical, as for very heavy particles \citep{srinivas2025weakly} and for the single-phase layer. Its magnitude depends much more on the flux and the density ratio than on the injection velocity. At $\mathcal{F}=0.5\mathcal{F}_0$, for instance, $\Re{a_1}$ is close to $-0.5$ for both heavy particle types and close to $-0.33$ for both light ones, and it hardly changes between $W^*/W_T=0.2$ and $1.2$. Raising the flux moves every curve towards zero. The saturated amplitude $A_e=\sqrt{-\mathcal{K}_c\mathcal{C}_i/\Re{a_1}}$ increases accordingly, so that, at the same distance from onset, a more heavily loaded layer convects more strongly once it has started.

At $\mathcal{F}=1.5\mathcal{F}_0$ the picture changes, and it does so differently for each density ratio. For the heaviest particles ($\beta=0.1$, figure~\ref{fig:Wstar vs a1r beta 0.1}), $\Re{a_1}$ stays negative, but it rises steeply as the injection velocity is reduced, from about $-0.45$ at $W^*/W_T=1.2$ to $-0.13$ at $W^*/W_T=0.2$. Slowly injected heavy particles crowd near the upper wall (figure~\ref{fig:heavy base-state}), and it is in this regime that the nonlinear saturation is weakest. For $\beta=0.5$ (figure~\ref{fig:Wstar vs a1r beta 0.5}) the same tendency is enough to change the sign: $\Re{a_1}$ is small and positive for sub-terminal injection and crosses zero at $W^*/W_T\approx1.01$, so the bifurcation is subcritical when the particles are injected below their terminal speed and supercritical when they are injected above it. For $\beta=2.5$ (figure~\ref{fig:Wstar vs a1r beta 2.5}) $\Re{a_1}$ is positive over the whole range of injection velocities, decreasing slowly from about $0.07$ to $0.01$. For $\beta=3$ (figure~\ref{fig:Wstar vs a1r beta 3.0}) the coefficient remains very close to zero and is positive only in a narrow window, $0.27\lesssim W^*/W_T\lesssim0.40$. This window lies next to the point where $Ra_c$ itself passes through zero (figure~\ref{fig:Wstar vs Rac beta 3.0}), that is, where the thermal buoyancy stops being the main driver of the instability, and the Archimedes force on the particles takes over.

We stress that, whenever $\Re{a_1}>0$, the cubic Landau equation predicts the existence of an unstable finite-amplitude branch below $Ra_c$ but not the amplitude at which the convection eventually saturates; for that, higher-order terms in the amplitude expansion are needed. The values of $\Re{a_1}$ in the subcritical cases of figure~\ref{fig:Wstar vs a1r} are also small, typically $0.01$--$0.07$, compared with $0.2$--$0.5$ in the supercritical cases. The subcritical branches are therefore weak, and the hysteresis they imply should be confined to a narrow range of Rayleigh numbers below $Ra_c$.

\subsection{Bifurcation map in the plane of density ratio and injection velocity}\label{sec:bifurcation map}
\begin{figure}
     \centering
     \begin{subfigure}[b]{0.49\textwidth}
         \centering
         \includegraphics[width=\textwidth]{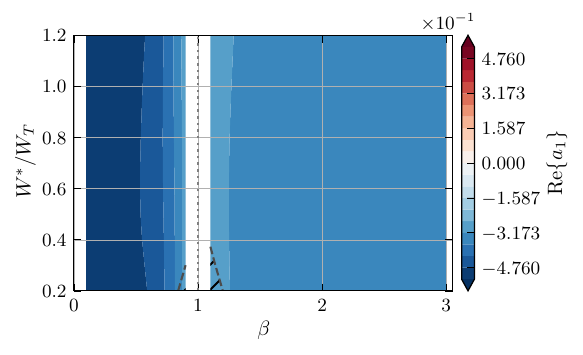}
         \caption{}
         \label{fig:plane a1r F0.5}
     \end{subfigure}
     \hfill
    \begin{subfigure}[b]{0.49\textwidth}
         \centering
         \includegraphics[width=\textwidth]{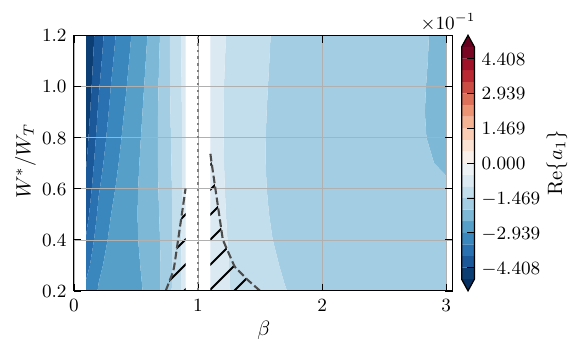}
         \caption{}
         \label{fig:plane a1r F1}
     \end{subfigure}
     \hfill
    \begin{subfigure}[b]{0.49\textwidth}
         \centering
         \includegraphics[width=\textwidth]{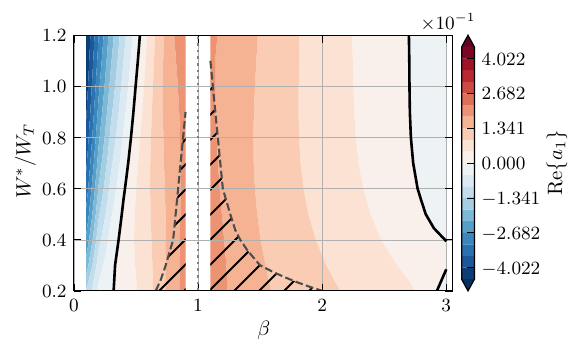}
         \caption{}
         \label{fig:plane a1r F1.5}
     \end{subfigure}
     \captionsetup{width=\linewidth}
    \caption{\justifying Variation in real part of Landau constant given by~\eqref{eq:a1} with the particle inlet velocity $W^*/W_T$ and density ratios $\beta$ at three different particle fluxes: (a) $\mathcal{F}=0.5\mathcal{F}_0$, (b) $\mathcal{F}=\mathcal{F}_0$ and (c) $\mathcal{F}=1.5\mathcal{F}_0$. Other parameters kept at $Ga=1.92\times10^{10}$, $Pr=5$, $\Phi=10^{-2}$, $E=5\times 10^{-3}$ and $\delta_{Ra}=5\times10^{-3}$. Here, the solid black line marks $\Re{a_1}=0$, the vertical dotted line marks neutrally buoyant particles ($\beta=1$, where $W_T=0$), and the hatched region marks an inlet particle volume fraction greater than $1\%$.}
    \label{fig:plane a1r}
\end{figure}

Figure~\ref{fig:plane a1r} extends these results to the whole $(\beta,\, W^*/W_T)$ plane. Neutrally buoyant particles, $\beta=1$, are excluded because their terminal velocity vanishes and the injection condition cannot be imposed; near this line the volume fraction needed to carry the flux becomes large, and the hatched regions mark where the inlet volume fraction exceeds one per cent. Results inside these regions should be treated with caution, since the dilute, point-particle description underlying the two-fluid model is no longer well justified there.

At $\mathcal{F}=0.5\mathcal{F}_0$ and $\mathcal{F}_0$ (figures~\ref{fig:plane a1r F0.5} and \ref{fig:plane a1r F1}) $\Re{a_1}$ is negative over the entire plane. The contours are almost vertical: the strength of the nonlinear saturation is set mainly by the density ratio, and is largest for the heaviest particles, while the injection velocity plays only a secondary role. The coefficient approaches zero as $\beta\to1$ from either side, which already hints at what happens when the loading is increased. At $\mathcal{F}=1.5\mathcal{F}_0$ (figure~\ref{fig:plane a1r F1.5}) a region of positive $\Re{a_1}$ appears. It is bounded on the left by a line that lies at $\beta\approx0.5$--$0.6$ and on the right by a line close to $\beta\approx2.6$--$3$, and it contains almost all of the light particles considered together with the moderately heavy ones. The very heavy particles on the left of the map remain supercritical, as do the lightest particles ($\beta\approx3$) when injected faster than about $0.4W_T$. Importantly, the subcritical region extends well beyond the hatched areas, so its existence does not rely on the parts of the map where the dilute approximation is questionable.

The two boundaries in figure~\ref{fig:plane a1r F1.5} are lines of degenerate, or tricritical, bifurcation points at which the cubic coefficient vanishes. On these lines, the type of the bifurcation is decided by the quintic term of the amplitude equation, which we have not computed. Taken together, figures~\ref{fig:Wstar vs a1r} and \ref{fig:plane a1r} show that the pRB system does not go from supercritical to subcritical simply because the particle density is reduced. Two conditions have to be met: the density contrast must be moderate, and the particle loading must be large enough. With these conditions met, the switch to subcritical onset occurs for heavy and light particles alike.

\subsection{Which balance equation decides the sign of $a_1$?}\label{sec:a1 split}
\begin{figure}
     \centering
     \begin{subfigure}[b]{0.49\textwidth}
         \centering
         \includegraphics[width=\textwidth]{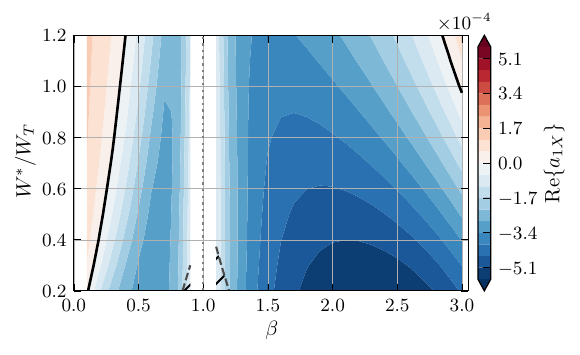}
         \caption{}
         \label{fig:split X F0.5}
     \end{subfigure}
     \hfill
    \begin{subfigure}[b]{0.49\textwidth}
         \centering
         \includegraphics[width=\textwidth]{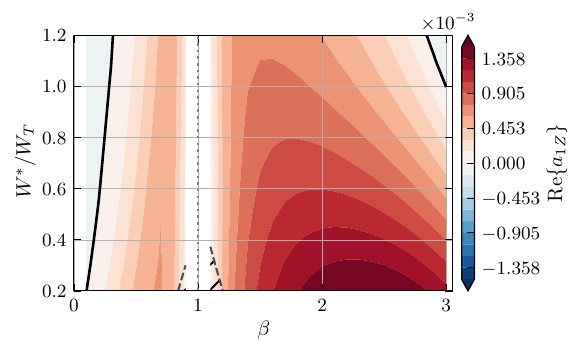}
         \caption{}
         \label{fig:split Z F0.5}
     \end{subfigure}
     \hfill
    \begin{subfigure}[b]{0.49\textwidth}
         \centering
         \includegraphics[width=\textwidth]{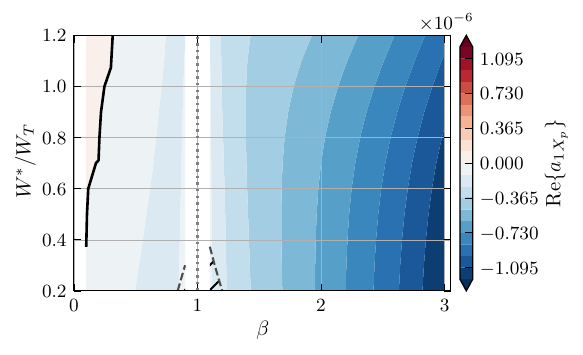}
         \caption{}
         \label{fig:split pX F0.5}
     \end{subfigure}
     \hfill
    \begin{subfigure}[b]{0.49\textwidth}
         \centering
         \includegraphics[width=\textwidth]{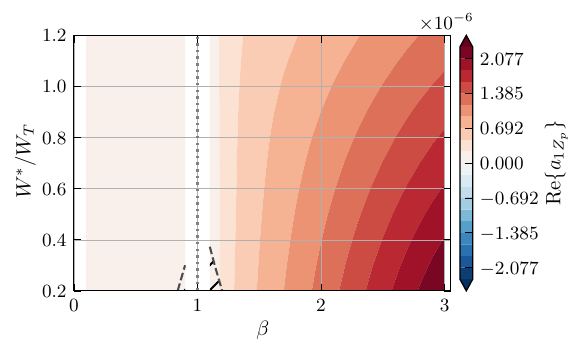}
         \caption{}
         \label{fig:split pZ F0.5}
     \end{subfigure}
    \hfill
    \begin{subfigure}[b]{0.49\textwidth}
         \centering
         \includegraphics[width=\textwidth]{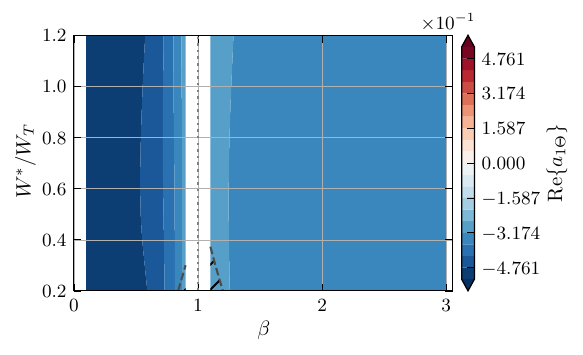}
         \caption{}
         \label{fig:split Theta F0.5}
     \end{subfigure}
    \hfill
    \begin{subfigure}[b]{0.49\textwidth}
         \centering
         \includegraphics[width=\textwidth]{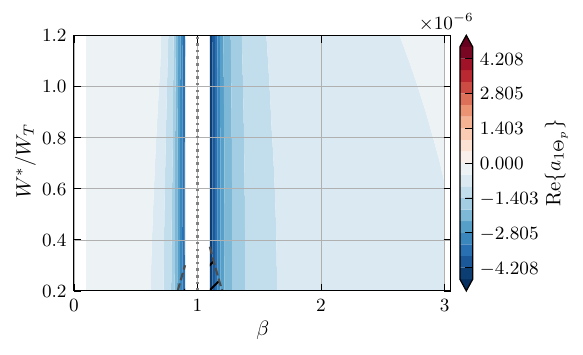}
         \caption{}
         \label{fig:split pTheta F0.5}
     \end{subfigure}
       \hfill
    \begin{subfigure}[b]{0.49\textwidth}
         \centering
         \includegraphics[width=\textwidth]{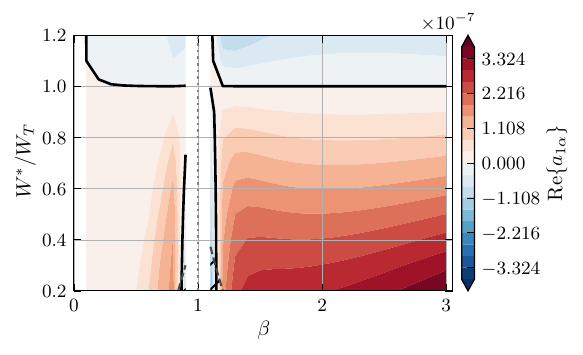}
         \caption{}
         \label{fig:split alpha F0.5}
     \end{subfigure}
     \captionsetup{width=\linewidth}
    \caption{\justifying Contributions of the individual terms in~\eqref{eq:a1} to the real part of the first Landau constant in the $\qty(\beta,\, W^*/W_T)$ plane for $\mathcal{F}=0.5\mathcal{F}_0$: (a) $\Re{a_{1X}}$, (b) $\Re{a_{1Z}}$, (c) $\Re{a_{1X_p}}$, (d) $\Re{a_{1Z_p}}$, (e) $\Re{a_{1\Theta}}$, (f) $\Re{a_{1\Theta_p}}$ and (g) $\Re{a_{1\alpha}}$, as defined in~\eqref{eq:a_1 terms}. Other parameters are $Ga = 1.92\times10^{10}$, $Pr = 5$, $\Phi = 10^{-2}$, $E = 5\times10^{-3}$ and $\delta_{Ra} = 5\times10^{-3}$. In each panel, the solid black line marks the zero contour of the plotted contribution, the vertical dotted line marks $\beta=1$, and the hatched region marks an inlet particle volume fraction greater than 1\%. Note the different colour scales.}
    \label{fig:split F0.5}
\end{figure}

\begin{figure}
     \centering
     \begin{subfigure}[b]{0.49\textwidth}
         \centering
         \includegraphics[width=\textwidth]{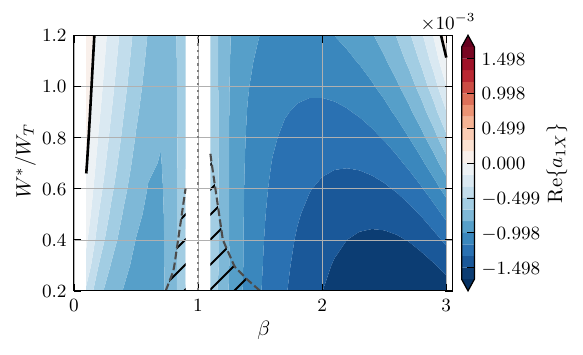}
         \caption{}
         \label{fig:split X F1}
     \end{subfigure}
     \hfill
    \begin{subfigure}[b]{0.49\textwidth}
         \centering
         \includegraphics[width=\textwidth]{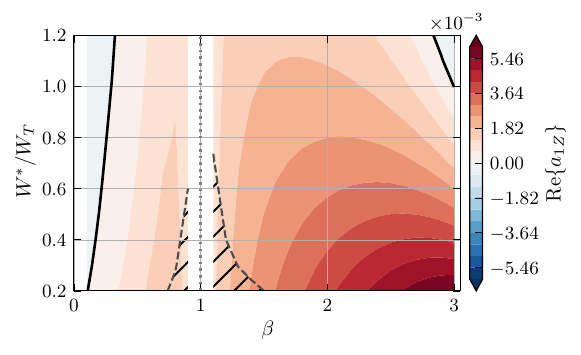}
         \caption{}
         \label{fig:split Z F1}
     \end{subfigure}
     \hfill
    \begin{subfigure}[b]{0.49\textwidth}
         \centering
         \includegraphics[width=\textwidth]{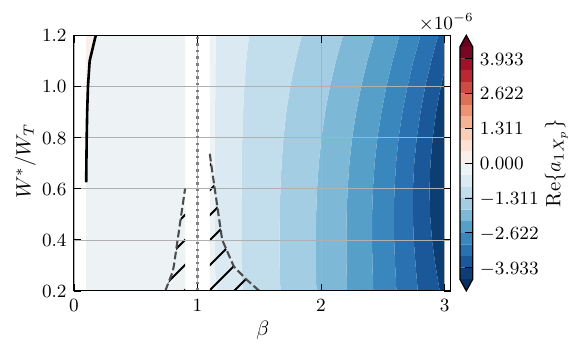}
         \caption{}
         \label{fig:split pX F1}
     \end{subfigure}
     \hfill
    \begin{subfigure}[b]{0.49\textwidth}
         \centering
         \includegraphics[width=\textwidth]{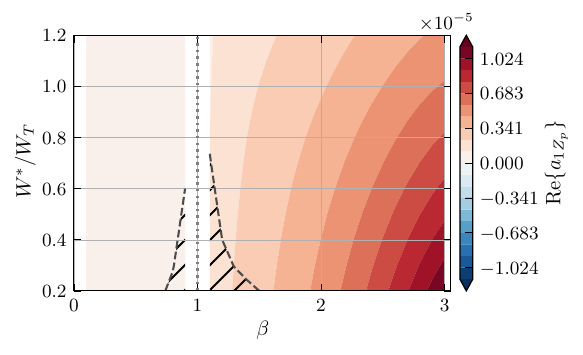}
         \caption{}
         \label{fig:split pZ F1}
     \end{subfigure}
    \hfill
    \begin{subfigure}[b]{0.49\textwidth}
         \centering
         \includegraphics[width=\textwidth]{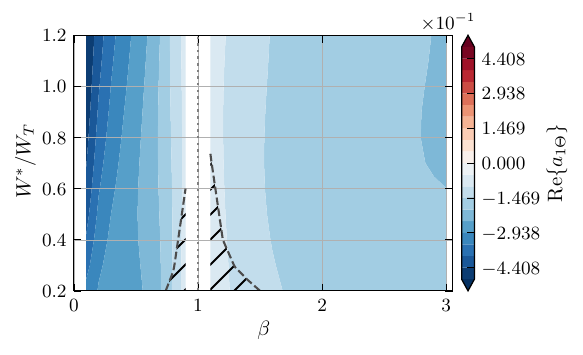}
         \caption{}
         \label{fig:split Theta F1}
     \end{subfigure}
    \hfill
    \begin{subfigure}[b]{0.49\textwidth}
         \centering
         \includegraphics[width=\textwidth]{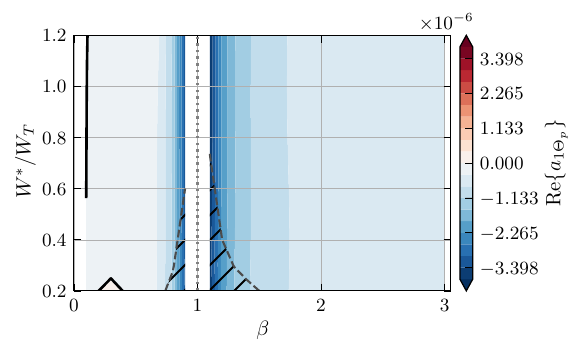}
         \caption{}
         \label{fig:split pTheta F1}
     \end{subfigure}
       \hfill
    \begin{subfigure}[b]{0.49\textwidth}
         \centering
         \includegraphics[width=\textwidth]{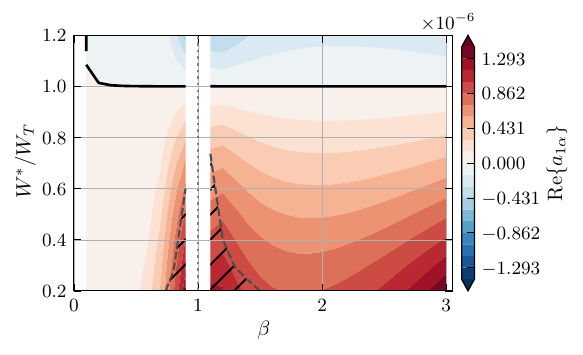}
         \caption{}
         \label{fig:split alpha F1}
     \end{subfigure}
     \captionsetup{width=\linewidth}
    \caption{\justifying Contributions of the individual terms in~\eqref{eq:a1} to the real part of the first Landau constant in the $\qty(\beta,\, W^*/W_T)$ plane for $\mathcal{F}=\mathcal{F}_0$: (a) $\Re{a_{1X}}$, (b) $\Re{a_{1Z}}$, (c) $\Re{a_{1X_p}}$, (d) $\Re{a_{1Z_p}}$, (e) $\Re{a_{1\Theta}}$, (f) $\Re{a_{1\Theta_p}}$ and (g) $\Re{a_{1\alpha}}$, as defined in~\eqref{eq:a_1 terms}. Other parameters are $Ga = 1.92\times10^{10}$, $Pr = 5$, $\Phi = 10^{-2}$, $E = 5\times10^{-3}$ and $\delta_{Ra} = 5\times10^{-3}$. In each panel, the solid black line marks the zero contour of the plotted contribution, the vertical dotted line marks $\beta=1$, and the hatched region marks an inlet particle volume fraction greater than 1\%. Note the different colour scales.}
    \label{fig:split F1}
\end{figure}

\begin{figure}
     \centering
     \begin{subfigure}[b]{0.49\textwidth}
         \centering
         \includegraphics[width=\textwidth]{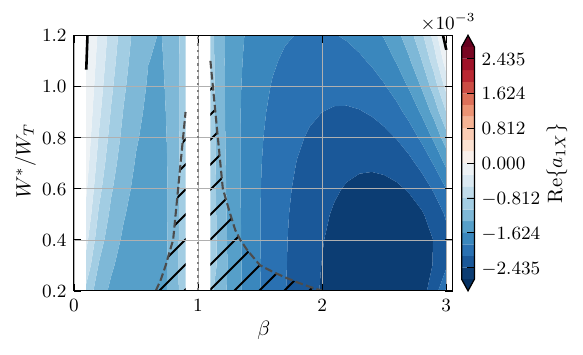}
         \caption{}
         \label{fig:split X F1.5}
     \end{subfigure}
     \hfill
    \begin{subfigure}[b]{0.49\textwidth}
         \centering
         \includegraphics[width=\textwidth]{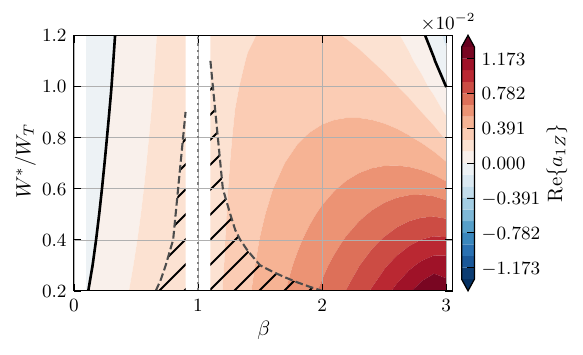}
         \caption{}
         \label{fig:split Z F1.5}
     \end{subfigure}
     \hfill
    \begin{subfigure}[b]{0.49\textwidth}
         \centering
         \includegraphics[width=\textwidth]{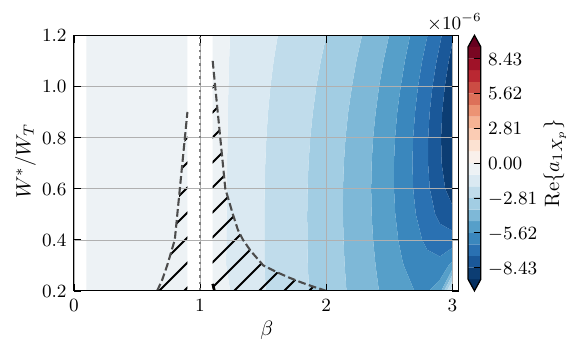}
         \caption{}
         \label{fig:split pX F1.5}
     \end{subfigure}
     \hfill
    \begin{subfigure}[b]{0.49\textwidth}
         \centering
         \includegraphics[width=\textwidth]{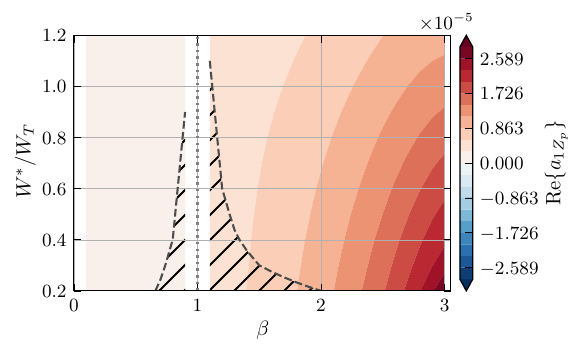}
         \caption{}
         \label{fig:split pZ F1.5}
     \end{subfigure}
    \hfill
    \begin{subfigure}[b]{0.49\textwidth}
         \centering
         \includegraphics[width=\textwidth]{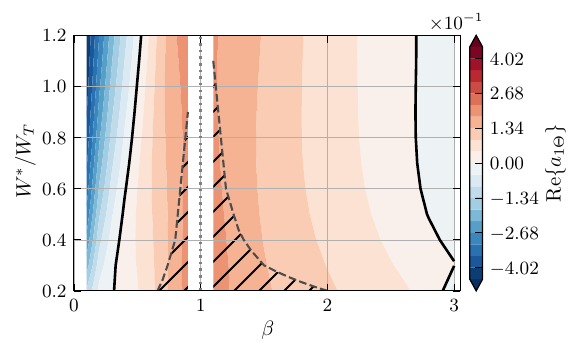}
         \caption{}
         \label{fig:split Theta F1.5}
     \end{subfigure}
    \hfill
    \begin{subfigure}[b]{0.49\textwidth}
         \centering
         \includegraphics[width=\textwidth]{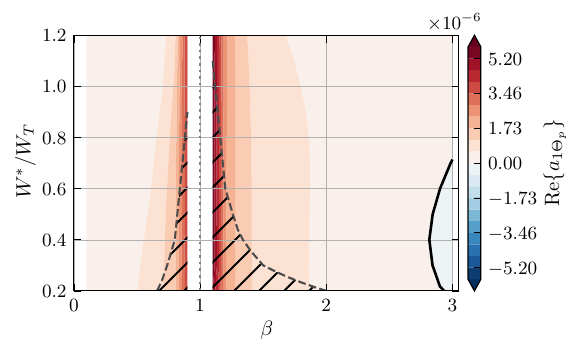}
         \caption{}
         \label{fig:split pTheta F1.5}
     \end{subfigure}
       \hfill
    \begin{subfigure}[b]{0.49\textwidth}
         \centering
         \includegraphics[width=\textwidth]{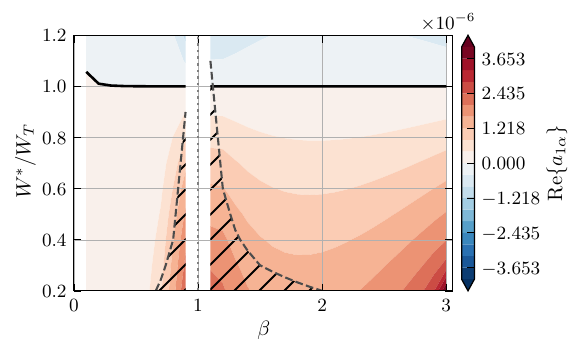}
         \caption{}
         \label{fig:split alpha F1.5}
     \end{subfigure}
     \captionsetup{width=\linewidth}
    \caption{\justifying Contributions of the individual terms in~\eqref{eq:a1} to the real part of the first Landau constant in the $\qty(\beta,\, W^*/W_T)$ plane for $\mathcal{F}=1.5\mathcal{F}_0$: (a) $\Re{a_{1X}}$, (b) $\Re{a_{1Z}}$, (c) $\Re{a_{1X_p}}$, (d) $\Re{a_{1Z_p}}$, (e) $\Re{a_{1\Theta}}$, (f) $\Re{a_{1\Theta_p}}$ and (g) $\Re{a_{1\alpha}}$, as defined in~\eqref{eq:a_1 terms}. Other parameters are $Ga = 1.92\times10^{10}$, $Pr = 5$, $\Phi = 10^{-2}$, $E = 5\times10^{-3}$ and $\delta_{Ra} = 5\times10^{-3}$. In each panel, the solid black line marks the zero contour of the plotted contribution, the vertical dotted line marks $\beta=1$, and the hatched region marks an inlet particle volume fraction greater than 1\%. Note the different colour scales.}
    \label{fig:split F1.5}
\end{figure}

The individual contributions defined in \eqref{eq:a_1 split} are shown in figures~\ref{fig:split F0.5}--\ref{fig:split F1.5}. Note that the colour scales differ by several orders of magnitude from panel to panel. The fluid energy contribution $a_{1\Theta}$ is of order $10^{-1}$ in every case, and its map reproduces that of $\Re{a_1}$ almost exactly, including the change of sign at $\mathcal{F}=1.5\mathcal{F}_0$ (figure~\ref{fig:split Theta F1.5}). The fluid momentum contributions are much smaller: $a_{1Z}$ is of order $10^{-3}$--$10^{-2}$ and is positive for light particles, so it pushes in the subcritical direction, whereas $a_{1X}$ is of order $10^{-4}$--$10^{-3}$ and mostly negative. The terms that originate in the particle equations are smaller still: $a_{1X_p}$, $a_{1Z_p}$ and $a_{1\Theta_p}$ are of order $10^{-6}$--$10^{-5}$, and $a_{1\alpha}$ of order $10^{-7}$--$10^{-6}$. They grow with the flux and are largest for light, slowly injected particles, but even there they are five to six orders of magnitude smaller than $a_{1\Theta}$.

At first sight, this seems to contradict the conclusion of \S~\ref{sec:bifurcation map}, where the particles were found to change the type of bifurcation. It does not, because each contribution to $a_1$ combines the nonlinear terms of one equation with the base state, with the fundamental mode, and with the mean distortion and second harmonic forced by it. The particles act on all of these. Through interphase heat exchange they reshape the base-state temperature profile; through drag and buoyancy they modify the structure of the linear eigenmode and of its adjoint; and through the coupling terms in \eqref{eq:E0 alpha}--\eqref{eq:E0 thetap} they alter the mean temperature distortion $\Theta_1$. In single-phase convection, the saturation of the rolls is due to this distortion, which flattens the mean temperature gradient in the interior and so reduces the buoyancy available to the mode \citep{malkus1958finite}. A plausible reading of figures~\ref{fig:split F0.5}--\ref{fig:split F1.5} is that, at high loading and moderate density contrast, the particles weaken or reverse this feedback, so that the convective heat transport no longer acts to limit its own growth. The direct nonlinear dynamics of the particle phase, by contrast, plays essentially no role close to onset. A detailed check of this interpretation would require the profiles of $\Theta_1$ and of the base-state temperature gradient for representative supercritical and subcritical cases, and we leave it to future work.

\section{Conclusions}\label{sec:conclusions}
We have examined the nature of the primary bifurcation in the particulate Rayleigh--B\'enard system, a fluid layer heated from below through which a dilute stream of thermally coupled particles is passed at a fixed flux. The particle phase was modelled with the two-fluid equations of \citet{raza2024stabilization, raza2025impact}. These retain drag, buoyancy, pressure-gradient, and added-mass forces and interphase heat exchange, and therefore cover particles that are much heavier than the fluid as well as bubbles. The linear stability results of \citet{raza2025impact} were reproduced, and a multiple-scale expansion with a Fredholm solvability condition was used to derive a cubic Landau equation for the amplitude of the convective rolls and to compute its coefficient over the $(\beta,\, W^*/W_T)$ plane for three particle fluxes.

The main findings are as follows.
\begin{enumerate}
\item At low and moderate loading, the bifurcation is supercritical for every density ratio and injection velocity considered. The conclusion of \citet{srinivas2025weakly} for very heavy particles thus carries over to light particles and bubbles, provided the flux is not too large.
\item Increasing the flux weakens the nonlinear saturation for all particle types. At $\mathcal{F}=1.5\mathcal{F}_0$ the real part of the Landau coefficient becomes positive, and the onset turns subcritical over a band of density ratios, $0.5\lesssim\beta\lesssim3$, that includes moderately heavy particles and most of the light particles studied. Very heavy particles remain supercritical, and so do the lightest particles when injected sufficiently fast.
\item Within this band, the injection velocity can decide the outcome. For $\beta=0.5$ the bifurcation is subcritical for sub-terminal injection and supercritical for super-terminal injection, with the transition close to $W^*=W_T$. For $\beta=3$, the subcritical window is narrow and lies next to the point where the critical Rayleigh number changes sign.
\item The sign of the Landau coefficient is decided almost entirely in the fluid energy equation. The contributions from the particle momentum, particle energy, and particle continuity equations are five to six orders of magnitude smaller. The particles therefore alter the bifurcation indirectly, by changing the base state, the linear mode and the mean temperature distortion, and not through their own nonlinear dynamics.
\end{enumerate}

These results answer, at least in part, the question that motivated this work: reducing the density of the particles relative to the fluid does turn the supercritical onset of particle-laden convection into a subcritical one, but only when the particle loading is sufficiently high. The subcritical branches found here are weak, which suggests that the associated hysteresis should be small, but this needs to be confirmed.

Several limitations of the present analysis point to natural extensions. The cubic Landau equation cannot predict the saturated amplitude of subcritical convection, nor the behaviour on the tricritical lines, so a quintic expansion, or direct numerical simulations of the two-fluid equations, are needed to quantify the hysteresis loop and the threshold amplitude below $Ra_c$. The analysis has been restricted to two-dimensional rolls. Because the particle injection breaks the up--down symmetry of the layer, quadratic interactions between three-dimensional modes are permitted, and hexagonal patterns may appear at onset in the same way as in non-Boussinesq convection \citep{busse1978non}; this deserves a separate study. Finally, near neutral buoyancy the required volume fractions exceed the range in which a dilute point-particle model can be trusted. That part of the parameter space would be better explored with a model that accounts for finite volume fraction effects, or with particle-resolved simulations such as those of \citet{chen2024particle}.




\begin{bmhead}[Declaration of interests]{The author reports no conflict of interest.}
\end{bmhead}


\begin{bmhead}[Author ORCID]{Thota Srinivas, https://orcid.org/0009-0004-0013-9157}
\end{bmhead}


\appendix
\begin{appen}
\section{Linear perturbation equations}\label{appen:linear perturbed system}
\begin{align}\label{eq:linear fluid mass balance}
     &\mathscr{L}_0\qty(\mathcal{K},\,\widehat{U}_x,\,\widehat{U}_z)=\text{i}\mathcal{K}\widehat{U}_x+D{\widehat{U}_z}=0,\\
     &\mathscr{L}_\alpha\qty(\mathcal{K},\, \mathcal{C},\, \widehat{\alpha},\, \widehat{W}_x,\, \widehat{W}_z,\, \alpha_0,\, W_0)=\text{i}\mathcal{C}\mathcal{K}\widehat{\alpha}-\alpha_0\qty(\text{i}\mathcal{K}\widehat{W}_x+D\widehat{W}_z)-\qty(DW_0)\,\widehat{\alpha}-W_0\,D\widehat{\alpha}\nonumber\\
     &\quad\quad-\qty(D\alpha_0)\,\widehat{W}_z=0,\\ \label{eq:linear Ux eq.}
     &\mathscr{L}_x\qty(\mathcal{K},\, \mathcal{C},\, \widehat{P},\, \widehat{U}_x,\, \widehat{W}_x,\, \alpha_0,\, \beta,\, Pr,\, \Phi)=\text{i}\mathcal{C}\mathcal{K}\qty(1-\frac{\alpha_0}{2}\qty(\beta-1))\widehat{U}_x+Pr\qty(D^2-\mathcal{K}^2)\widehat{U}_x-\text{i}\mathcal{K}\widehat{P} \nonumber\\
     &\quad-6Pr\qty(3-\beta)\Phi^{-2}\alpha_0\qty(\widehat{U}_x-\widehat{W}_x)=0,\\ \label{eq:linear Uz eq.}
    &\mathscr{L}_z\qty(\mathcal{K},\, \mathcal{C},\, \widehat{P},\, \widehat{\alpha},\, \widehat{U}_z,\, \widehat{W}_z,\, \widehat{\Theta},\, \alpha_0,\, W_0,\, \beta,\, Ra,\, Pr,\, Ga,\, \Phi)=\text{i}\mathcal{C}\mathcal{K}\qty(1-\frac{\alpha_0}{2}\qty(\beta-1))\widehat{U}_z\nonumber\\&\quad+Pr\qty(D^2-\mathcal{K}^2)\widehat{U}_z-D\widehat{P}
     -6Pr\qty(3-\beta)\Phi^{-2}\alpha_0\qty(\widehat{U}_z-\widehat{W}_z)+6Pr(3-\beta)W_0\Phi^{-2}\widehat{\alpha}\nonumber\\
     &\quad\quad+\frac{\qty(\beta-1)GaPr^2}{2}\widehat{\alpha}+RaPr\widehat{\Theta}=0,\\
     &\mathscr{L}_{px}\qty(\mathcal{K},\, \mathcal{C},\, \widehat{U}_x,\, \widehat{W}_x,\, W_0,\, \beta,\, Pr, \Phi)=\text{i}\mathcal{C}\mathcal{K}\qty(\widehat{W}_x-\beta\widehat{U}_x)-W_0\,D\widehat{W}_x+12Pr\beta\Phi^{-2}\qty(\widehat{U}_x-\widehat{W}_x)=0,\\
    &\mathscr{L}_{pz}\qty(\mathcal{K},\, \mathcal{C},\, \widehat{U}_z,\, \widehat{W}_z,\, W_0,\, \beta,\, Pr, \Phi)=\text{i}\mathcal{C}\mathcal{K}\qty(\widehat{W}_z-\beta\widehat{U}_z)-W_0\,D\widehat{W}_z-\qty(DW_0)\widehat{W}_z\nonumber\\
    &\quad+12Pr\beta\Phi^{-2}\qty(\widehat{U}_z-\widehat{W}_z)=0,\\
    &\mathscr{L}_{\theta}\qty(\mathcal{K},\, \mathcal{C},\, \widehat{\alpha},\, \widehat{U}_z,\, \widehat{\Theta},\, \widehat{\Theta}_{p},\, \alpha_0,\, \Theta_0,\, \Theta_{p0},\, \Phi)=\text{i}\mathcal{C}\mathcal{K}\qty(1-\alpha_0)\widehat{\Theta}-\qty(1-\alpha_0)\qty(D\Theta_0)\widehat{U}_z+\qty(D^2-\mathcal{K}^2)\widehat{\Theta}\nonumber\\
    &\quad-12\alpha_0\Phi^{-2}\qty(\widehat{\Theta}-\widehat{\Theta}_p)-12\Phi^{-2}\qty(\Theta_0-\Theta_{p0})\widehat{\alpha}=0,\\ \label{eq:linear particle energy balance}
    &\mathscr{L}_{p\theta}\qty(\mathcal{K},\, \mathcal{C},\, \widehat{W}_z,\, \widehat{\Theta},\, \widehat{\Theta}_p,\, E,\ Pr,\, \Phi)=\text{i}\mathcal{C}\mathcal{K}\widehat{\Theta}_p-W_0\,D\widehat{\Theta}_p-\qty(D\Theta_{p0})\widehat{W}_z+12E^{-1}\Phi^{-2}\qty(\widehat{\Theta}-\widehat{\Theta}_p)=0
\end{align}
for $Z\in\qty(-1/2,\,1/2)$ and the boundary conditions are given by (\ref{eq:linear BCs}).

\section{Linear energy budget equations}\label{appen:linear KE equations}
By eliminating the pressure from (\ref{eq:linear Ux eq.}) and (\ref{eq:linear Uz eq.}), and using (\ref{eq:linear fluid mass balance}) we obtain the following equation
\begin{equation}
    \begin{split}
        \text{i}\mathcal{C}\mathcal{K}\qty{\qty(1-\frac{\alpha_0}{2}\qty(\beta-1))\qty(D^2-\mathcal{K}^2)\widehat{U}_z -\frac{\beta-1}{2}\qty(D\alpha_0)\qty(D\widehat{U}_z)  }=-Pr\qty(D^2-\mathcal{K}^2)^2\widehat{U}_z\\
        +PrRa\mathcal{K}^2\widehat{\Theta}+6Pr\qty(3-\beta)\Phi^{-2}\qty(D\alpha_0)\qty(D\widehat{U}_z+\text{i}\mathcal{K}\widehat{W}_x)\\
        +6Pr\qty(3-\beta)\Phi^{-2}\alpha_0\qty(\text{i}\mathcal{K}D\widehat{W}_x+\qty(D^2-\mathcal{K}^2)\widehat{U}_z+\mathcal{K}^2\widehat{W}_z)\\
        +\qty(\frac{\beta-1}{2}GaPr^2 + 6Pr\qty(3-\beta)\Phi^{-2}W_0)\mathcal{K}^2\alpha.
    \end{split}
\end{equation}
Now, by multiplying by the complex conjugate of $\widehat{U}_z$ represented as $\widetilde{U}_z$, and integrating over the whole domain, we get the following energy budget equation,
\begin{equation}\label{eq:linear energy balance}
    \mathcal{C}_i\mathcal{K}e^K = e^\Theta+e^V+e^{\alpha_0W}+e^{\alpha W_0} + e^{\alpha Ga},
\end{equation}
where each term is defined as
\begin{align}
    e^K&=-\int_{Z=-1/2}^{1/2} \Re{ \qty[\qty(1-\frac{\alpha_0}{2}\qty(\beta-1))\qty(D^2-\mathcal{K}^2)\widehat{U}_z -\frac{\beta-1}{2}\qty(D\alpha_0)\qty(D\widehat{U}_z)  ] \widetilde{U}_z} \,dZ,\\
    e^\Theta&=RaPr\int_{Z=-1/2}^{1/2} \Re{\mathcal{K}^2\widehat{\Theta}\widetilde{U}_z} \,dZ,\\
    e^V&=-Pr\int_{Z=-1/2}^{1/2} \Re{\qty[\qty(D^2-\mathcal{K}^2)^2\widehat{U}_z] \widetilde{U}_z} \,dZ,\\
    e^{\alpha_0W}&=6Pr\qty(3-\beta)\Phi^{-2}\left[\int_{Z=-1/2}^{1/2} \Re{\qty[\alpha_0\qty(\text{i}\mathcal{K}D\widehat{W}_x+\qty(D^2-\mathcal{K}^2)\widehat{U}_z+\mathcal{K}^2\widehat{W}_z)]\widetilde{U}_z} \,dZ \right.\nonumber\\
    &\quad+\left. \int_{Z=-1/2}^{1/2} \Re{\qty[\qty(D\alpha_0)\qty(D\widehat{U}_z+\text{i}\mathcal{K}\widehat{W}_x)]\widetilde{U}_z} \,dZ \right],\\
    e^{\alpha W_0}&=6Pr\qty(3-\beta)\Phi^{-2}\int_{Z=-1/2}^{1/2} \Re{\qty[W_0\mathcal{K}^2\widehat{\alpha}] \widetilde{U}_z} \,dZ,\\
    e^{\alpha Ga}&=\frac{\beta-1}{2}GaPr^2\int_{Z=-1/2}^{1/2} \Re{\qty[\mathcal{K}^2\widehat{\alpha}]\widetilde{U}_z} \,dZ,
\end{align}
where the term $e^K>0$ because its leading-order contribution is given by
\begin{equation}
    \int_{Z=-1/2}^{1/2} \Re{\qty[\qty(D^2-\mathcal{K}^2)\widehat{U}_z]\widetilde{U}_z}\,dZ=-\int_{Z=-1/2}^{1/2}\qty(\abs{D\widehat{U}_z}^2+\mathcal{K}^2\abs{\widehat{U}_z}^2)\,dZ<0.
\end{equation}
Following~\cite{raza2025impact}, the physical interpretation of various terms in (\ref{eq:linear energy balance}) is as follows: $e^V$ represents the rate of viscous dissipation, $e^\Theta$ is the power of the thermal buoyancy force, and $e^{\alpha_0W}$, $e^{\alpha W_0}$ and $e^{\alpha Ga}$ correspond to the two-way coupling between the fluid and particle phases. Particularly, $e^{\alpha_0W}$ and $e^{\alpha W_0}$ are the rates of work done by particles due to non-uniform base-state particle volume fraction and settling velocity, respectively. The term $e^{\alpha Ga}$ represents the rate of work done by Archimedes' force.

\section{Equations at different harmonics}\label{appen:harmonic equations}
Equations for the harmonic $\mathbb{E}^0$ are given by
\begin{align}\label{eq:E0 alpha}
    &\mathcal{L}_\alpha(\alpha_0,\, W_0)+\mathcal{C}_i\abs{B}^2\qty{D\qty(\alpha_1W_0+\alpha_0W_{z1}) }+\mathcal{C}_i\abs{B}^2\qty{D\qty(\widetilde{\alpha}_{10}W_{z10})+D\qty(\alpha_{10}\widetilde{W}_{z10})}=\order{\mathcal{C}_i^2},\\
    &\mathcal{L}_z\qty(\alpha_0,\, P_0,\, W_0,\, \Theta_0,\, Ra,\, Pr,\, Ga,\, \beta,\, \Phi)\nonumber\\
    &\quad+\mathcal{C}_i\abs{B}^2\qty{\mathcal{L}_z\qty(\alpha_1,\, P_1,\, W_0,\, \Theta_1,\, Ra,\, Pr,\, Ga,\, \beta,\, \Phi)-6Pr\qty(3-\beta)\Phi^{-2}\alpha_0W_{z1}}\nonumber\\
    &\quad\quad+\mathcal{C}_i\abs{B}^2\left\{ \frac{1}{2}\text{i}\mathcal{K}_c\qty(\beta-1)\qty(\mathcal{C}\widetilde{\alpha}_{10}U_{z10}-\widetilde{\mathcal{C}}\alpha_{10}\widetilde{U}_{z10})\right.\\
    &\quad\quad\quad\left.+\qty(1-\frac{\alpha_0}{2}\qty(\beta-1))\qty(\text{i}\mathcal{K}_c\widetilde{U}_{x10}U_{z10}-\text{i}\mathcal{K}_cU_{x10}\widetilde{U}_{z10}+D\qty(\widetilde{U}_{z10}U_{z10}))\right.\nonumber\\
    &\quad\quad\quad\quad\left.+6Pr\qty(3-\beta)\Phi^{-2}\qty(\widetilde{\alpha}_{10}\qty(U_{z10}-W_{z10})+\alpha_{10}\qty(\widetilde{U}_{z10}-\widetilde{W}_{z10})) \right\}=\order{\mathcal{C}_i^2},\\
    &\mathcal{L}_{pz}\qty(W_0,\, Ga,\, Pr,\, \beta,\, \Phi)+\mathcal{C}_i\abs{B}^2\qty{W_0\qty(DW_{z1})+W_{z1}\qty(DW_{0})+12Pr\beta\Phi^{-2}W_{z1}}\nonumber\\
    &\quad+\mathcal{C}_i\abs{B}^2\left\{\text{i}\mathcal{K}_c\widetilde{W}_{x10}W_{z10}-\text{i}\mathcal{K}_cW_{x10}\widetilde{W}_{z10}+\widetilde{W}_{z10}\qty(DW_{z10})+W_{z10}\qty(D\widetilde{W}_{z10})\right.\nonumber\\
    &\quad\quad \left.-\beta\qty(\text{i}\mathcal{K}_c\widetilde{U}_{x10}U_{z10}-\text{i}\mathcal{K}_cU_{x10}\widetilde{U}_{z10}+D\qty(U_{z10}\widetilde{U}_{z10})) \right\}=\order{\mathcal{C}_i^2},\\
    &\mathcal{L}_\theta\qty(\alpha_0,\, \Theta_0,\, \Theta_{p0},\, \Phi)+\mathcal{C}_i\abs{B}^2\qty{\mathcal{L}_\theta\qty(\alpha_0,\, \Theta_1,\, \Theta_{p1},\, \Phi)-12\Phi^{-2}\alpha_1\qty(\Theta_0-\Theta_{p0})}\nonumber\\
    &\quad+\mathcal{C}_i\abs{B}^2\left\{\text{i}\mathcal{K}_c\qty(\widetilde{\mathcal{C}}\alpha_{10}\widetilde{\Theta}_{10}-\mathcal{C}\widetilde{\alpha}_{10}\Theta_{10})-12\Phi^{-2}\qty[\alpha_{10}\qty(\widetilde{\Theta}_{10}-\widetilde{\Theta}_{p10})+\widetilde{\alpha}_{10}\qty(\Theta_{10}-\Theta_{p10})] \right.\nonumber\\
    &\quad\quad\left. +\text{i}\mathcal{K}_c\qty(1-\alpha_0)\qty(U_{x10}\widetilde{\Theta}_{10}-
    \widetilde{U}_{x10}\Theta_{10})-\qty(1-\alpha_0)\qty(\widetilde{U}_{z10}\qty(D\Theta_{10})+U_{z10}\qty(D\widetilde{\Theta}_{10})) \right.\nonumber\\
    &\quad\quad\quad\left. + \qty(D\Theta_0)\qty(U_{z10}\widetilde{\alpha}_{10}+\widetilde{U}_{z10}\alpha_{10})\right\}=\order{\mathcal{C}_i^2},\\ \label{eq:E0 thetap}
    &\mathcal{L}_{p\theta}\qty(W_0,\, \Theta_0,\, \Theta_{p0},\, E,\, \Phi)+\mathcal{C}_i\abs{B}^2\qty{\mathcal{L}_{p\theta}\qty(W_0,\, \Theta_1,\, \Theta_{p1},\, E,\, \Phi)+W_{z1}\qty(D\Theta_{p0})}\nonumber\\
    &\quad+\mathcal{C}_i\abs{B}^2\qty{\text{i}\mathcal{K}_c\widetilde{W}_{x10}\Theta_{p10}-\text{i}\mathcal{K}_cW_{x10}\widetilde{\Theta}_{p10}+\widetilde{W}_{z10}\qty(D\Theta_{p10})+W_{z10}\qty(D\widetilde{\Theta}_{p10})}=\order{\mathcal{C}_i^2}
\end{align}
for $Z\in\qty(-1/2,\, 1/2)$ and the operators $\mathcal{L}_\alpha,\, \mathcal{L}_z,\, \mathcal{L}_{pz},\, \mathcal{L}_\theta$ and $\mathcal{L}_{p\theta}$ are given, respectively by (\ref{eq:base state alpha})--(\ref{eq:basestate thetap equation}). The boundary conditions for $\alpha_1,\, W_{z1},\, \Theta_1$ and $\Theta_{p1}$ are given by
\begin{equation}\label{eq:base state distortion BCs}
    \left.\begin{array}{l}
    \mbox{At }Z=1/2:\,\Theta_1=0, \\[12pt]
    \mbox{At }Z=-1/2:\,\Theta_1=0,\\[12pt]
    \mbox{At }Z=Z^*:\,{W}_{z1}=0,\, \Theta_{p1}=0, \mbox{ and } \alpha_1=0.
    \end{array}\right\}
\end{equation}

Similarly, equations for the harmonic $\mathbb{E}^1$ are given by
\begin{align}\label{eq:E1 fluid mass balance}
    &\qty(\mathcal{C}_i)^{1/2}B\mathscr{L}_0\qty(\mathcal{K}_c,\, U_{x10},\, U_{z10})+\qty(\mathcal{C}_i)^{3/2}B\abs{B}^2\mathscr{L}_0\qty(\mathcal{K}_c,\, U_{x11},\, U_{z11})=\order{\mathcal{C}_i^{5/2}},\\ \label{eq:E1 alpha}
    &\qty(\mathcal{C}_i)^{1/2}B\mathscr{L}_\alpha\qty(\mathcal{K}_c,\, \mathcal{C},\, \alpha_{10},\, W_{x10},\, W_{z10},\, \alpha_0,\, W_0)\nonumber\\
    &\quad+\qty(\mathcal{C}_i)^{3/2}B\abs{B}^2\mathscr{L}_\alpha\qty(\mathcal{K}_c,\, \mathcal{C},\, \alpha_{11},\, W_{x11},\, W_{z11},\, \alpha_0,\, W_0)\nonumber\\
    &\quad\quad-\qty(\mathcal{C}_i^{3/2})\qty(B\abs{B}^2\mathcal{G}_\alpha+\qty{\dv{B}{\tau}-\mathcal{K}_cB}\alpha_{10})=\order{\mathcal{C}_i^{5/2}},\\
    &\qty(\mathcal{C}_i)^{1/2}B\mathscr{L}_x\qty(\mathcal{K}_c,\, \mathcal{C},\, P_{10},\, U_{x10},\, W_{x10},\, \alpha_0,\, \beta,\, Pr,\, \Phi)\nonumber\\
    &\quad+\qty(\mathcal{C}_i)^{3/2}B\abs{B}^2\mathscr{L}_x\qty(\mathcal{K}_c,\, \mathcal{C},\, P_{11},\, U_{x11},\, W_{x11},\, \alpha_0,\, \beta,\, Pr,\, \Phi)\nonumber\\
    &\quad\quad-\qty(\mathcal{C}_i^{3/2})\qty(B\abs{B}^2\mathcal{G}_x+\qty{\dv{B}{\tau}-\mathcal{K}_cB}\qty(1-\frac{\alpha_0}{2}\qty(\beta-1))U_{x10})=\order{\mathcal{C}_i^{5/2}},\\
    &\qty(\mathcal{C}_i)^{1/2}B\mathscr{L}_z\qty(\mathcal{K}_c,\, \mathcal{C},\, P_{10},\, \alpha_{10},\, U_{z10},\, W_{z10},\, \Theta_{10},\, \alpha_0,\, W_0,\, \beta,\, Ra,\, Pr,\, Ga,\, \Phi)\nonumber\\
    &\quad+\qty(\mathcal{C}_i)^{3/2}B\abs{B}^2\mathscr{L}_z\qty(\mathcal{K}_c,\, \mathcal{C},\, P_{11},\, \alpha_{11},\, U_{z11},\, W_{z11},\, \Theta_{11},\, \alpha_0,\, W_0,\, \beta,\, Ra,\, Pr,\, Ga,\, \Phi)\nonumber\\
    &\quad\quad-\qty(\mathcal{C}_i^{3/2})\qty(B\abs{B}^2\mathcal{G}_z+\qty{\dv{B}{\tau}-\mathcal{K}_cB}\qty(1-\frac{\alpha_0}{2}\qty(\beta-1))U_{z10})=\order{\mathcal{C}_i^{5/2}},\\
    &\qty(\mathcal{C}_i)^{1/2}B\mathscr{L}_{px}\qty(\mathcal{K}_c,\, \mathcal{C},\, U_{x10},\, W_{x10},\, W_0,\, \beta,\, Pr, \Phi)\nonumber\\
    &\quad+\qty(\mathcal{C}_i)^{3/2}B\abs{B}^2\mathscr{L}_{px}\qty(\mathcal{K}_c,\, \mathcal{C},\, U_{x11},\, W_{x11},\, W_0,\, \beta,\, Pr, \Phi)\nonumber\\
    &\quad\quad-\qty(\mathcal{C}_i^{3/2})\qty(B\abs{B}^2\mathcal{G}_{px}+\qty{\dv{B}{\tau}-\mathcal{K}_cB}\qty(W_{x10}-\beta U_{x10}))=\order{\mathcal{C}_i^{5/2}},\\
    &\qty(\mathcal{C}_i)^{1/2}B\mathscr{L}_{pz}\qty(\mathcal{K}_c,\, \mathcal{C},\, U_{z10},\, W_{z10},\, W_0,\, \beta,\, Pr, \Phi)\nonumber\\
    &\quad+\qty(\mathcal{C}_i)^{3/2}B\abs{B}^2\mathscr{L}_{pz}\qty(\mathcal{K}_c,\, \mathcal{C},\, U_{z11},\, W_{z11},\, W_0,\, \beta,\, Pr, \Phi)\nonumber\\
     &\quad\quad-\qty(\mathcal{C}_i^{3/2})\qty(B\abs{B}^2\mathcal{G}_{pz}+\qty{\dv{B}{\tau}-\mathcal{K}_cB}\qty(W_{z10}-\beta U_{z10}))=\order{\mathcal{C}_i^{5/2}},\\
     &\qty(\mathcal{C}_i)^{1/2}B\mathscr{L}_{\theta}\qty(\mathcal{K}_c,\, \mathcal{C},\, \alpha_{10},\, U_{z10},\, \Theta_{10},\, \Theta_{p10},\, \alpha_0,\, \Theta_0,\, \Theta_{p0},\, \Phi)\nonumber\\
     &\quad+\qty(\mathcal{C}_i)^{3/2}B\abs{B}^2\mathscr{L}_{\theta}\qty(\mathcal{K}_c,\, \mathcal{C},\, \alpha_{11},\, U_{z11},\, \Theta_{11},\, \Theta_{p11},\, \alpha_0,\, \Theta_0,\, \Theta_{p0},\, \Phi)\nonumber\\
     &\quad\quad-\qty(\mathcal{C}_i^{3/2})\qty(B\abs{B}^2\mathcal{G}_{\theta}+\qty{\dv{B}{\tau}-\mathcal{K}_cB}\qty(1-\alpha_0)\Theta_{10})=\order{\mathcal{C}_i^{5/2}},\\ \label{eq:E1 thetap}
     &\qty(\mathcal{C}_i)^{1/2}B \mathscr{L}_{p\theta}\qty(\mathcal{K}_c,\, \mathcal{C},\, W_{z10},\, \Theta_{10},\, \Theta_{p10},\, E,\ Pr,\, \Phi)\nonumber\\
     &\quad+\qty(\mathcal{C}_i)^{3/2}B\abs{B}^2\mathscr{L}_{p\theta}\qty(\mathcal{K}_c,\, \mathcal{C},\, W_{z11},\, \Theta_{11},\, \Theta_{p11},\, E,\ Pr,\, \Phi)\nonumber\\
     &\quad\quad-\qty(\mathcal{C}_i^{3/2})\qty(B\abs{B}^2\mathcal{G}_{p\theta}+\qty{\dv{B}{\tau}-\mathcal{K}_cB}\Theta_{p10})=\order{\mathcal{C}_i^{5/2}}
\end{align}
for $Z\in\qty(-1/2,\, 1/2)$ and the linear operators $\mathscr{L}_0,\, \mathscr{L}_\alpha,\, \mathscr{L}_x,\, \mathscr{L}_z,\, \mathscr{L}_{px},\, \mathscr{L}_{pz},\, \mathscr{L}_\theta$ and $\mathscr{L}_{p\theta}$ are given, respectively by (\ref{eq:linear fluid mass balance})--(\ref{eq:linear particle energy balance}), and both $\qty(\alpha_{10},\, U_{x10},\, U_{z10},\, W_{x10},\, W_{z10},\, \Theta_{10},\, \Theta_{p10})$  and $\qty(\alpha_{11},\, U_{x11},\, U_{z11},\, W_{x11},\, W_{z11},\, \Theta_{11},\, \Theta_{p11})$ are subjected to similar boundary conditions given by (\ref{eq:linear BCs}). The scalar functions $\qty(\mathcal{G}_\alpha,\, \mathcal{G}_x,\, \mathcal{G}_z,\, \mathcal{G}_{px},\, \mathcal{G}_{pz},\, \mathcal{G}_{\theta},\, \mathcal{G}_{p\theta})$ are given by~(\ref{eq:G_alpha})--(\ref{eq:G_ptheta}) in \S~\ref{appen:scalar functions}.

Finally, equations at harmonic $\mathbb{E}^2$ are given by
\begin{align}\label{eq:E2 mass balance}
    &\mathcal{C}_iB^2\mathscr{L}_0\qty(2\mathcal{K}_c,\, U_{x20},\, U_{z20})=\order{\mathcal{C}_i^2},\\
    &\mathcal{C}_iB^2\mathscr{L}_\alpha\qty(2\mathcal{K}_c,\, \mathcal{C},\, \alpha_{20},\, W_{x20},\, W_{z20},\, \alpha_0,\, W_0)\nonumber\\
    &\quad-\mathcal{C}_iB^2\qty{2\text{i}\mathcal{K}_c\alpha_{10}W_{x10}+D\qty(\alpha_{10}W_{z10})}=\order{\mathcal{C}_i^2}, \\
    &\mathcal{C}_iB^2\mathscr{L}_x\qty(2\mathcal{K}_c,\, \mathcal{C},\, P_{20},\, U_{x20},\, W_{x20},\, \alpha_0,\, \beta,\, Pr,\, \Phi)\nonumber\\
    &\quad-\mathcal{C}_iB^2\left\{ \frac{\text{i}}{2}\qty(\beta-1)\mathcal{C}\mathcal{K}_c\alpha_{10}U_{x10}+\qty(1-\frac{\alpha_0}{2}\qty(\beta-1))\qty[\text{i}\mathcal{K}_cU_{x10}^2+U_{z10}\qty(DU_{x10})]\right.\nonumber\\
    &\quad\quad\left.+6Pr\qty(3-\beta)\Phi^{-2}\alpha_{10}\qty(U_{x10}-W_{x10})\right\}=\order{\mathcal{C}_i^2},\\
    &\mathcal{C}_iB^2\mathscr{L}_z\qty(2\mathcal{K}_c,\, \mathcal{C},\, P_{20},\, \alpha_{20},\, U_{z20},\, W_{z20},\, \Theta_{20},\, \alpha_0,\, W_0,\, \beta,\, Ra,\, Pr,\, Ga,\, \Phi)\nonumber\\
    &\quad-\mathcal{C}_iB^2\left\{ \frac{\text{i}}{2}\qty(\beta-1)\mathcal{C}\mathcal{K}_c\alpha_{10}U_{z10}+\qty(1-\frac{\alpha_0}{2}\qty(\beta-1))\qty[\text{i}\mathcal{K}_cU_{x10}U_{z10}+U_{z10}\qty(DU_{z10})]\right.\nonumber\\
    &\quad\quad\left.+6Pr\qty(3-\beta)\Phi^{-2}\alpha_{10}\qty(U_{z10}-W_{z10})\right\}=\order{\mathcal{C}_i^2},\\
    &\mathcal{C}_iB^2\mathscr{L}_{px}\qty(2\mathcal{K}_c,\, \mathcal{C},\, U_{x20},\, W_{x20},\, W_0,\, \beta,\, Pr, \Phi)\nonumber\\
    &\quad-\mathcal{C}_iB^2\qty{\text{i}\mathcal{K}_c\qty(W_{x10}^2-\beta U_{x10}^2)+W_{z10}\qty(DW_{x10})-\beta U_{z10}\qty(DU_{x10})}=\order{\mathcal{C}_i^2},\\
    &\mathcal{C}_iB^2\mathscr{L}_{pz}\qty(2\mathcal{K}_c,\, \mathcal{C},\, U_{z20},\, W_{z20},\, W_0,\, \beta,\, Pr, \Phi)\nonumber\\
    &\quad-\mathcal{C}_iB^2\qty{\text{i}\mathcal{K}_c\qty(W_{x10}W_{z10}-\beta U_{x10}U_{z10})+W_{z10}\qty(DW_{z10})-\beta U_{z10}\qty(DU_{z10})}=\order{\mathcal{C}_i^2},\\
    &\mathcal{C}_iB^2\mathscr{L}_{\theta}\qty(2\mathcal{K}_c,\, \mathcal{C},\, \alpha_{20},\, U_{z20},\, \Theta_{20},\, \Theta_{p20},\, \alpha_0,\, \Theta_0,\, \Theta_{p0},\, \Phi)\nonumber\\
    &\quad-\mathcal{C}_iB^2\left\{\text{i}\mathcal{C}\mathcal{K}_c\alpha_{10}\Theta_{10}+\text{i}\mathcal{K}_c\qty(1-\alpha_0)U_{x10}\Theta_{10}+\qty(1-\alpha_0)U_{z10}\qty(D\Theta_{10})-\qty(D\Theta_0)\alpha_{10}U_{z10}\right.\nonumber\\
    &\quad\quad\left.+12\Phi^{-2}\alpha_{10}\qty(\Theta_{10}-\Theta_{p10})\right\}=\order{\mathcal{C}_i^2},\\ \label{eq:E2 thetap}
    &\mathcal{C}_iB^2\mathscr{L}_{p\theta}\qty(2\mathcal{K}_c,\, \mathcal{C},\, W_{z20},\, \Theta_{20},\, \Theta_{p20},\, E,\ Pr,\, \Phi)\nonumber\\
    &\quad-\mathcal{C}_iB^2\qty{\text{i}\mathcal{K}_cW_{x10}\Theta_{p10}+W_{z10}\qty(D\Theta_{p10})}=\order{\mathcal{C}_i^2}
\end{align}
for $Z\in\qty(-1/2,\, 1/2)$ and the boundary conditions are similar to (\ref{eq:linear BCs}).

\section{Scalar equations}\label{appen:scalar functions}
\begin{align}\label{eq:G_alpha}
    \mathcal{G}_\alpha&=\text{i}\mathcal{K}_c\widetilde{\alpha}_{10}W_{x20}+\text{i}\mathcal{K}_c\alpha_{20}\widetilde{W}_{x10}+\text{i}\mathcal{K}_c\alpha_1W_{x10}\nonumber\\
    &\quad+\alpha_{10}DW_{z1}+\alpha_1DW_{z10}+\alpha_{20}D\widetilde{W}_{z10}+\widetilde{\alpha}_{10}DW_{z20}+\qty(D\alpha_1)W_{z10}+\qty(D\alpha_{10})W_{z1}\nonumber\\
    &\quad\quad+\qty(D\widetilde{\alpha}_{10})W_{z20}+\qty(D\alpha_{20})\widetilde{W}_{z10},\\
    \mathcal{G}_x&=\text{i}\mathcal{K}_c\qty(\beta-1)\qty(-\frac{1}{2}\widetilde{\mathcal{C}}\alpha_{20}\widetilde{U}_{x10}+\mathcal{C}\widetilde{\alpha}_{10}U_{x20}+\frac{1}{2}\mathcal{C}\alpha_1U_{x10})+2\text{i}\mathcal{K}_c\widetilde{U}_{x10}U_{x20}-\text{i}\mathcal{K}_c\widetilde{U}_{x10}U_{x20}\nonumber\\
    &\quad+\qty(D\widetilde{U}_{x10})U_{z20}+\qty(DU_{x20})\widetilde{U}_{z10}\nonumber\\
    &\quad\quad+6Pr\qty(3-\beta)\Phi^{-2}\qty(\widetilde{\alpha}_{10}\qty(U_{x20}-W_{x20})+\alpha_1\qty(U_{x10}-W_{x10})+\alpha_{20}\qty(\widetilde{U}_{x10}-\widetilde{W}_{x10}))\nonumber\\
    &\quad\quad\quad-\frac{\beta-1}{2}\qty(\text{i}\mathcal{K}_c\widetilde{\alpha}_{10}U_{x10}^2+\text{i}\mathcal{K}_c\alpha_0\widetilde{U}_{x10}U_{x20})\nonumber\\
    &\quad\quad\quad\quad-\frac{\beta-1}{2}\qty(\alpha_0\qty{\qty(D\widetilde{U}_{x10})U_{z20}+\qty(DU_{x20})\widetilde{U}_{z10}})\nonumber\\
    &\quad\quad\quad\quad\quad-\frac{\beta-1}{2}\qty(\alpha_{10}\qty(DU_{x10})\widetilde{U}_{z10}+\alpha_{10}\qty(D\widetilde{U}_{x10}){U}_{z10}+\widetilde{\alpha}_{10}\qty(DU_{x10})U_{z10}),\\
    \mathcal{G}_z&=\text{i}\mathcal{K}_c\qty(\beta-1)\qty(-\frac{1}{2}\widetilde{\mathcal{C}}\alpha_{20}\widetilde{U}_{z10}+\mathcal{C}\widetilde{\alpha}_{10}U_{z20}+\frac{1}{2}\mathcal{C}\alpha_1U_{z10})+2\text{i}\mathcal{K}_c\widetilde{U}_{x10}U_{z20}-\text{i}\mathcal{K}_c\widetilde{U}_{z10}U_{x20}\nonumber\\
    &\quad+\qty(D\widetilde{U}_{z10})U_{z20}+\qty(DU_{z20})\widetilde{U}_{z10}\nonumber\\
    &\quad\quad+6Pr\qty(3-\beta)\Phi^{-2}\qty(\widetilde{\alpha}_{10}\qty(U_{z20}-W_{z20})+\alpha_1\qty(U_{z10}-W_{z10})-\alpha_{10}W_{z1}+\alpha_{20}\qty(\widetilde{U}_{z10}-\widetilde{W}_{z10}))\nonumber\\
    &\quad\quad\quad-\frac{\beta-1}{2}\qty(\text{i}\mathcal{K}_c\widetilde{\alpha}_{10}U_{x10}U_{z10}+2\text{i}\mathcal{K}_c\alpha_0\widetilde{U}_{x10}U_{z20}+\text{i}\mathcal{K}_c\alpha_{10}\widetilde{U}_{x10}U_{z10})\nonumber\\
    &\quad\quad\quad\quad+\frac{\beta-1}{2}\qty(\text{i}\mathcal{K}_c\alpha_0U_{x20}\widetilde{U}_{z10}+\text{i}\mathcal{K}_c\alpha_{10}U_{x10}\widetilde{U}_{z10})\nonumber\\
    &\quad\quad\quad\quad\quad-\frac{\beta-1}{2}\qty(\alpha_0\qty{\qty(D\widetilde{U}_{z10})U_{z20}+\qty(DU_{z20})\widetilde{U}_{z10}})\nonumber\\
    &\quad\quad\quad\quad\quad\quad-\frac{\beta-1}{2}\qty(\alpha_{10}\qty(DU_{z10})\widetilde{U}_{z10}+\alpha_{10}\qty(D\widetilde{U}_{z10}){U}_{z10}+\widetilde{\alpha}_{10}\qty(DU_{z10})U_{z10}),\\
    \mathcal{G}_{px}&=\text{i}\mathcal{K}_c\widetilde{W}_{x10}W_{x20}+\qty(DW_{x10})W_{z1}+\qty(D\widetilde{W}_{x10})W_{z20}+\qty(DW_{x20})\widetilde{W}_{z10}\nonumber\\
    &\quad-\beta\qty{2\text{i}\mathcal{K}_c\widetilde{U}_{x10}U_{x20}-\text{i}\mathcal{K}_c\widetilde{U}_{x10}U_{x20}+\qty(D\widetilde{U}_{x10})U_{z20}+\qty(DU_{x20})\widetilde{U}_{z10}},\\
    \mathcal{G}_{pz}&=2\text{i}\mathcal{K}_c\widetilde{W}_{x10}W_{z20}-\text{i}\mathcal{K}_c\widetilde{W}_{z10}W_{x20}+\qty(DW_{z1})W_{z10}+\qty(DW_{z10})W_{z1}\nonumber\\
    &\quad+\qty(D\widetilde{W}_{z10})W_{z20}+\qty(DW_{z20})\widetilde{W}_{z10}-\beta\qty(2\text{i}\mathcal{K}_c\widetilde{U}_{x10}U_{z20}-\text{i}\mathcal{K}_c\widetilde{U}_{z10}U_{x20})\nonumber\\
    &\quad\quad-\beta\qty(\qty(D\widetilde{U}_{z10})U_{z20}+\qty(DU_{z20})\widetilde{U}_{z10}),\\
    \mathcal{G}_\theta&=\text{i}\mathcal{K}_c\qty(\mathcal{C}\alpha_1\Theta_{10}+2\mathcal{C}\widetilde{\alpha}_{10}\Theta_{20}-\widetilde{\mathcal{C}}\alpha_{20}\widetilde{\Theta}_{10})+2\text{i}\mathcal{K}_c\widetilde{U}_{x10}\Theta_{20}-\text{i}\mathcal{K}_cU_{x20}\widetilde{\Theta}_{10}\nonumber\\
    &\quad-\text{i}\mathcal{K}_c\alpha_0\qty(2\widetilde{U}_{x10}\Theta_{20}-U_{x20}\widetilde{\Theta}_{10})-\text{i}\mathcal{K}_c\widetilde{\alpha}_{10}U_{x10}\Theta_{10}-\text{i}\mathcal{K}_c\alpha_{10}\qty(\widetilde{U}_{x10}\Theta_{10}-U_{x10}\widetilde{\Theta}_{10})\nonumber\\
    &\quad\quad+\qty(D\Theta_1)U_{z10}+\qty(D\widetilde{\Theta}_{10})U_{z20}+\qty(D\Theta_{20})\widetilde{U}_{z10}-\qty(D\Theta_1)\alpha_0U_{z10}-\qty(D\Theta_{20})\alpha_0\widetilde{U}_{z10}\nonumber\\
    &\quad\quad\quad-\qty(D\Theta_0)\qty(\widetilde{\alpha}_{10}U_{z20}+\alpha_1U_{z10}+\alpha_{20}\widetilde{U}_{z10})-\qty(D\widetilde{\Theta}_{10})\qty(\alpha_{10}U_{z10}+\alpha_0U_{z20})\nonumber\\
    &\quad\quad\quad\quad-\qty(D\Theta_{10})\qty(\widetilde{\alpha}_{10}U_{z10}+\alpha_{10}\widetilde{U}_{z10})+12\Phi^{-2}\qty(\alpha_{10}\qty(\Theta_1-\Theta_{p1})+\alpha_1\qty(\Theta_{10}-\Theta_{p10}))\nonumber\\
    &\quad\quad\quad\quad\quad+12\Phi^{-2}\widetilde{\alpha}_{10}\qty(\Theta_{20}-\Theta_{p20})+12\Phi^{-2}\alpha_{20}\qty(\widetilde{\Theta_{10}}-\widetilde{\Theta_{p10}}),\\ \label{eq:G_ptheta}
    \mathcal{G}_{p\theta}&=2\text{i}\mathcal{K}_c\widetilde{W}_{x10}\Theta_{p20}-\text{i}\mathcal{K}_cW_{x20}\widetilde{\Theta}_{p10}+\qty(D\Theta_{p1})W_{z10}+\qty(D\Theta_{p10})W_{z1}\nonumber\\
    &\quad+\qty(D\widetilde{\Theta}_{p10})W_{z20}+\qty(D\Theta_{p20})\widetilde{W}_{z10}.
\end{align}

\section{Adjoint linear equations}\label{appen:adjoint linear system}
The adjoint of a linear operator $\mathscr{L}$ is represented by $\mathscr{L}^\dag$ and is defined as
\begin{equation}\label{eq:adjoint_def}
    \langle\boldsymbol{X}^\dag,\, \mathscr{L}\boldsymbol{X} \rangle = \langle \mathscr{L}^\dag\boldsymbol{X}^\dag,\, \boldsymbol{X}\rangle,
\end{equation}
where $\boldsymbol{X}^\dag$ represents the eigenfunction of adjoint operator $\mathscr{L}^\dag$. Here, the inner product between two complex-valued vector functions
$\boldsymbol{X} = [x_1,\, x_2,\, x_3,\, x_4,\, x_5,\, x_6,\, x_7,\, x_8]^T$ and
$\boldsymbol{Y} = [y_1,\, y_2,\, y_3,\, y_4,\, y_5,\, y_6,\, y_7,\, y_8]^T$ is defined as the
\emph{bilinear} (unconjugated) form
\begin{equation}
\langle \boldsymbol{X}, \boldsymbol{Y} \rangle = \int_{Z=-1/2}^{1/2} [\boldsymbol{X}]^T [\boldsymbol{Y}] \, dZ
= \int_{Z=-1/2}^{1/2} \sum_{j=1}^{8} x_j(Z) \, y_j(Z) \, dZ .
\label{eq:inner_product}
\end{equation}
This form, rather than the Hermitian one, is the appropriate pairing here: it is the bilinear form for which the adjoint operator $\mathscr{L}^{\dagger}$ defined by \eqref{eq:adjoint_def} possesses the same eigenvalue $\mathcal{C}$ as $\mathscr{L}$, so that the Fredholm solvability condition applied to \eqref{eq:E1 alpha}--\eqref{eq:E1 thetap} yields \eqref{eq:a1}. Using the above definition for the adjoint operator, the adjoint linear equations corresponding to the system of linear equations given by (\ref{eq:linear fluid mass balance})--(\ref{eq:linear particle energy balance}) are given by
\begin{align}
    &\mathscr{L}_0^\dag\qty(\mathcal{K}_c,\, U^\dag_x,\, U^\dag_z)=-\text{i}\mathcal{K}_cU^\dag_x+DU^\dag_z=0,\\
    &\mathscr{L}_\alpha^\dag\qty(\mathcal{K}_c,\, \mathcal{C},\, \alpha^\dag,\, U_z^\dag,\, W_0,\, Ga,\, Pr,\, \beta,\,  \Phi)=\text{i}\mathcal{C}\mathcal{K}_c\alpha^\dag+W_0\qty(D\alpha^\dag)+6Pr\qty(3-\beta)W_0\Phi^{-2}U_z^\dag\nonumber\\
    &\quad+\frac{\beta-1}{2}GaPr^2U_z^\dag-12\Phi^{-2}\qty(\Theta_0-\Theta_{p0})\Theta^\dag=0,\\
    &\mathscr{L}_x^\dag\qty(\mathcal{K}_c,\, \mathcal{C},\, P^\dag,\, U^\dag_x,\, W_x^\dag, \alpha_0,\, Pr,\beta,\, \Phi)=\text{i}\mathcal{C}\mathcal{K}_c\qty{\qty(1-\frac{\alpha_0}{2}\qty(\beta-1))U_x^\dag-\beta W_x^\dag}\nonumber\\
    &\quad+Pr\qty(D^2-\mathcal{K}_c^2)U_x^\dag-6Pr\qty(3-\beta)\Phi^{-2}\alpha_0U_x^\dag+\text{i}\mathcal{K}_cP^\dag+12Pr\beta\Phi^{-2}W_{x}^\dag=0,\\
    &\mathscr{L}_z^\dag\qty(\mathcal{K}_c,\, \mathcal{C},\, P^\dag,\, U^\dag_z,\, W_z^\dag,\, \Theta^\dag,\, \alpha_0,\,  \Theta_0,\, Pr,\, \beta,\, \Phi)=\text{i}\mathcal{C}\mathcal{K}_c\qty{\qty(1-\frac{\alpha_0}{2}\qty(\beta-1))U_z^\dag-\beta W_z^\dag}\nonumber\\
    &\quad+Pr\qty(D^2-\mathcal{K}_c^2)U_z^\dag-6Pr\qty(3-\beta)\Phi^{-2}\alpha_0U_z^\dag-DP^\dag\nonumber\\
    &\quad\quad+12Pr\beta\Phi^{-2}W_z^\dag-\qty(1-\alpha_0)\qty(D\Theta_0)\Theta^\dag=0,\\
    &\mathscr{L}_{px}^\dag\qty(\mathcal{K}_c,\, \mathcal{C},\, \alpha^\dag,\, U_{x}^\dag,\, W_x^\dag,\, \alpha_0,\, Pr,\, \beta,\, \Phi)=\text{i}\mathcal{C}\mathcal{K}_cW_x^\dag+D\qty(W_0W_x^\dag)-12Pr\beta\Phi^{-2}W_x^\dag\nonumber\\
    &\quad-\text{i}\mathcal{K}_c\alpha_0\alpha^\dag+6Pr\qty(3-\beta)\Phi^{-2}\alpha_0U_x^\dag=0,\\
    &\mathscr{L}_{pz}^\dag\qty(\mathcal{K}_c,\, \mathcal{C},\, \alpha^\dag,\, U_z^\dag,\, W_z^\dag,\, \Theta_p^\dag,\, \alpha_0,\, Pr,\, \beta,\, \Phi)=\text{i}\mathcal{C}\mathcal{K}_cW_z^\dag+W_0\qty(DW_z^\dag)-12Pr\beta\Phi^{-2}W_z^\dag\nonumber\\
    &\quad+\alpha_0\qty(D\alpha^\dag)+6Pr\qty(3-\beta)\Phi^{-2}\alpha_0U_z^\dag-\qty(D\Theta_{p0})\Theta_p^\dag=0,\\
    &\mathscr{L}_\theta^\dag\qty(\mathcal{K}_c,\, \mathcal{C},\, U_z^\dag,\, \Theta^\dag,\, \Theta_p^\dag,\, \alpha_0,\, Ra,\, Pr,\, E,\, \Phi)=\text{i}\mathcal{C}\mathcal{K}_c\qty(1-\alpha_0)\Theta^\dag+\qty(D^2-\mathcal{K}_c^2)\Theta^\dag\nonumber\\
    &\quad-12\alpha_0\Phi^{-2}\Theta^\dag+RaPrU_z^\dag+12E^{-1}\Phi^{-2}\Theta_p^\dag=0,\\
    &\mathscr{L}_{p\theta}^\dag\qty(\mathcal{K}_c,\, \mathcal{C},\, \Theta^\dag,\, \Theta_p^\dag,\, \alpha_0,\, W_0,\, E,\, \Phi)=\text{i}\mathcal{C}\mathcal{K}_c\Theta_p^\dag+D\qty(W_0\Theta_p^\dag)-12E^{-1}\Phi^{-2}\Theta_p^\dag\nonumber\\
    &\quad+12\alpha_0\Phi^{-2}\Theta^\dag=0,
\end{align}
for $Z\in\qty(-1/2,\, 1/2)$ and the boundary conditions are given by
\begin{equation}\label{eq:adjoing linear BCs}
    \left.\begin{array}{l}
    \mbox{At }Z=1/2:\,U^\dag_x=U^\dag_z=0 \mbox{ and } \Theta^\dag=0, \\[12pt]
    \mbox{At }Z=-1/2:\,U^\dag_x=U^\dag_z=0 \mbox{ and } \Theta^\dag=0,\\[12pt]
    \mbox{At }Z=-Z^*:\,W^\dag_x=W^\dag_z=0,\, \Theta^\dag_{p}=0, \mbox{ and } \alpha^\dag=0.
    \end{array}\right\}
\end{equation}
The adjoint eigenfunctions are normalized such that
\begin{align}\label{eq:adjoint normalisation}
    &\int_{Z=-1/2}^{1/2}\left\{\alpha_{10}\alpha^\dag+\qty(1-\frac{\alpha_0}{2}\qty(\beta-1))\qty(U_{x10}U_x^\dag+U_{z10}U_z^\dag)+\qty(W_{x10}-\beta U_{x10})W_x^\dag\right.\nonumber\\
    &\quad\quad\quad\quad\left.+\qty(W_{z10}-\beta U_{z10})W_z^\dag + \qty(1-\alpha_0)\Theta_{10}\Theta^\dag+\Theta_{p10}\Theta_p^\dag\right\}\,dZ=1.
\end{align}
\end{appen}

\bibliographystyle{jfm}
\bibliography{jfm}
\end{document}